\documentclass[twocolumn,aps,prb,showpacs]{revtex4-1}
\usepackage{amsmath,amssymb}
\usepackage{graphicx}
\usepackage{dcolumn}
\usepackage{bm}
\usepackage{hyperref}
\usepackage{xcolor}
\usepackage{caption}
\usepackage{subcaption}

\def\avg#1{{\langle #1\rangle }}
\def\kB{{k_{\rm B}}}
\def\rmd{{\rm d}}
\def\br{{\bf r}}

\begin{document}

\title{Quasicrystal structure prediction: A review}
\author{Michael Widom}
\affiliation{
  Physics Department, Carnegie Mellon University.
}
\author{Marek Mihalkovi\v c }
\affiliation{
  Inst. of Physics, Slovak Academy of Sciences, 84511 Bratislava, Slovakia
}

\date{\today}

\begin{abstract}

Predicting quasicrystal structures is a multifaceted problem that can involve predicting a previously unknown phase, predicting the structure of an experimentally observed phase, or predicting the thermodynamic stability of a given structure.  We survey the history and current state of these prediction efforts with a focus on methods that have improved our understanding of the structure and stability of known metallic quasicrystal phases. Advances in the structural modeling of quasicrystals, along with first principles total energy calculation and statistical mechanical methods that enable the calculation of quasicrystal thermodynamic stability, are illustrated by means of cited examples of recent work.
\end{abstract}
\maketitle

\section{Introduction}

This review surveys the past and present state of quasicrystal structure prediction. We address several aspects of this problem. First, we consider how the very existence of a quasicrystal could be anticipated, in the absence of prior experimental observation. Second, restricting our attention to metallic alloys, we address methods to predict the structure of a specific quasicrystalline phase given basic information concerning composition and lattice parameters. The problem is complicated by the likelihood that the quasicrystal is only stable at high T, hence, we finally address the problem of free energy calculation including various entropy contributions. We frame our discussion in the context of several specific examples from our own work and many others.

\section{Predicting the existence of quasicrystal phases}

Consider the problem of predicting formation of a quasicrystal in a novel alloy or soft-matter system; how can one proceed? The discovery by Penrose~\cite{Penrose1974,Gardner1977} of quasiperiodic tilings generalized the notion of a crystal lattice~\cite{Mackay1982} but provided no guidance for how to form such a structure in the laboratory. By chance, Shechtman and co-workers~\cite{Shechtman1984} discovered icosahedral symmetry in an Aluminum-rich alloy, Al$_{86}$Mn$_{14}$, and its quasiperiodicity was quickly identified~\cite{Levine1984}. Given that initial hint, within a year additional quasicrystals were discovered by chemical analogy in Al$_{86}$Cr$_{14}$ and Al$_{86}$Fe$_{14}$~\cite{Dunlap1985}, stepping one column of the periodic table to the left and right, respectively, from Mn. One can also move from binary to ternary through averaging of the valence electron count, hence replacing Mn with a Cr-Fe mixture also yields a quasicrystal~\cite{Dubois1986}. With the multiplicative effect of additional elements increasing the size of the composition space~\cite{Widom2017}, such substitution practices greatly increased the number of known quasicrystal-forming alloy systems, with the majority remaining Al-rich. One may question whether those successes represent {\em prediction} of quasicrystal formation, as opposed to {\em educated guesswork}.

Another approach, also based on guesswork, could be to mine existing knowledge of phase diagrams for unknown structures. In the case of Al-Cu-Fe, the $\Psi$-phase of Al$_6$Cu$_2$Fe, identified in 1939 with unknown structure~\cite{Bradley1939}, was likely the icosahedral quasicrystal. The same is true of the T$_2$ phase of Al-Cu-Li~\cite{hardy1955,Dubost1986}.  Similarly, the Cd$_{5.7}$Yb\cite{palenzona71} alloy possessed an unknown phase that turned out to be an example of the first thermodynamically stable binary quasicrystal\cite{tsai2000}. In practice, the approach can be more systematic, by searching for solved structures that contain quasicrystalline motifs, known as approximants~\cite{eh85}, hoping that a genuine quasicrystal may lie nearby. Examples of this approach include the search for boron-based icosahedral phases~\cite{KimuraB}, and the discovery of Frank-Kasper-type quasicrystals~\cite{FK}. The logical extension of this strategy is to enhance the search with methods of machine learning. Models can be trained on existing knowledge then new predictions can be tested experimentally~\cite{KimuraML}.

True ab-initio prediction would mean choosing an alloy system based on chemical knowledge or intuition, curiosity, or random guess, then searching the space of structures and (genetic algorithms, basin hopping, Monte Carlo, etc.) comparing energies or free energies with known existing phases. This has not been done to-date for a real quasicrystal; the closest example is decagonal B-Ti-Ru~\cite{KimuraBTiRu} which was discovered experimentally following a chemical substitution on a fully first principles prediction~\cite{BMgRu}.

Several artificial interatomic interaction models are known to produce quasicrystals, such as binary Lennard-Jones systems~\cite{Lancon,BinaryLJ}, double-well potentials~\cite{Dzugutov,Kiselev}, and polyhedral hard cores~\cite{Haji2011}. The extreme case is to specifically design a system to encourage a certain symmetry, or even to force quasiperiodicity. This has been achieved in colloidal systems using patchy interactions~\cite{Doye} and may be possible using designed proteins or DNA molecules in the future~\cite{DNA}.

An emerging area is the creation of artificial quasicrystals through the superposition of density waves. Early examples include the hydrodynamic patterns arising from Faraday instabilities in shaken fluids~\cite{edwards_fauve_1994}. Quasiperiodic patterns can be generated either by superposing density waves at angles whose tangents are irrational (e.g. with a crystallographically forbidden symmetry) or by forcing waves of incommensurate frequencies~\cite{LifshitzPetrich,Rutledge}. Recently optical quasilattices have been used to trap cold atoms in quasicrystalline arrays~\cite{Mace,ColdAtoms} and electrons have been subjected to quasiperiodic moir\'{e} potentials in twisted trilayer graphene~\cite{delaBarrera}. These last cases achieve seemingly perfect quasiperiodicity with near certainty, as the experiment itself directly follows a mathematical prescription to create quasiperiodicity.

\section{Predicting the structures of known quasicrystal phases}

Following the initial quasicrystal discovery, multiple structural models were quickly proposed. These included generalized Penrose tilings~\cite{Levine1984}, that could explain quasiperiodicity but lacked atomistic detail, and the icosahedral glass model~\cite{ShechtmanBlech85,StephensGoldman} that explained the propagation of orientational order but failed to explain quasiperiodicity. Linus Pauling~\cite{Pauling85,Pauling87} famously, but incorrectly, proposed models for decagonal and icosahedral quasicrystals based on multiple twinning of cubic crystals.

\begin{figure}
  \includegraphics[width=0.48\textwidth]{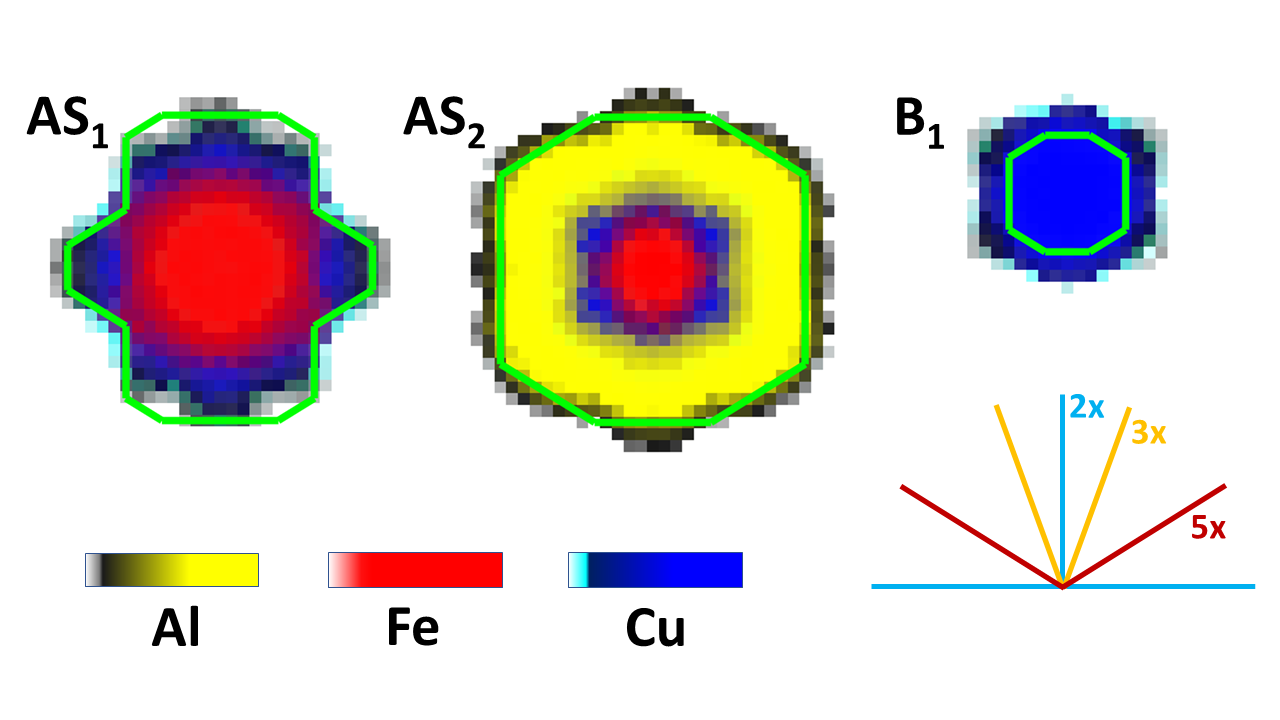}
  \includegraphics[width=0.2\textwidth]{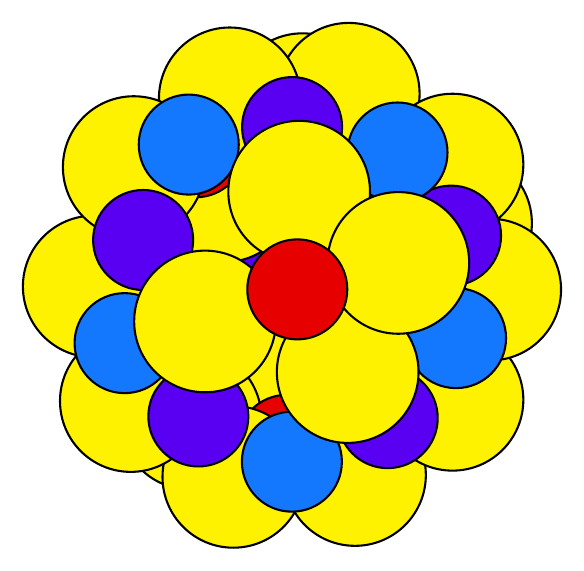}
  \includegraphics[width=0.25\textwidth]{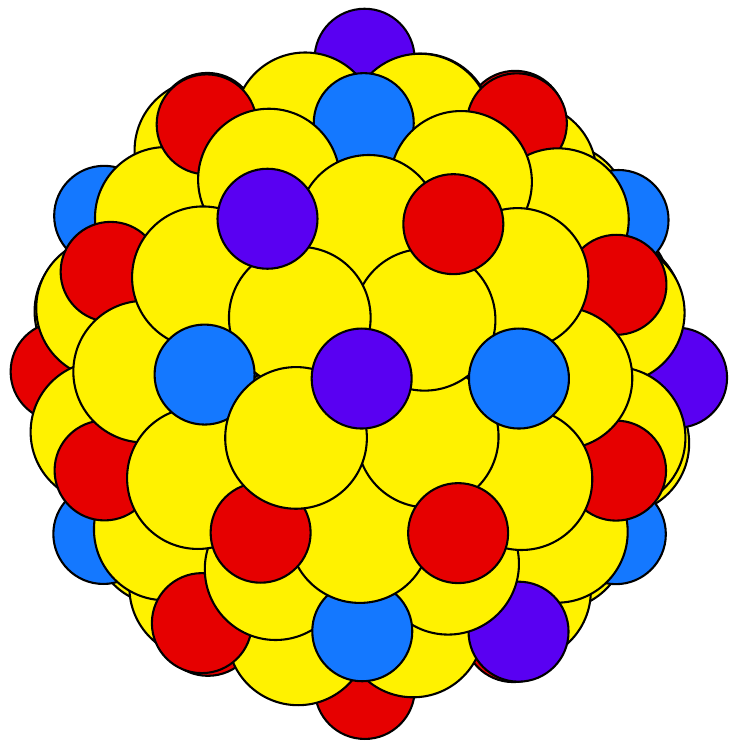}
  \caption{\label{fig:perp}
    (top) Simulated atomic surface occupation for $i$-AlCuFe~\cite{i-AlCuFe} at T=1242K; color bars for chemical species occupancy. Mixed chemical occupation is represented by adding the RGB color values. Green lines are Katz-Gratias atomic surface boundaries~\cite{kg93}. (bottom) pseudo-Mackay and $\tau$-scaled pseudo-Mackay icosahedral clusters showing sites of mixed Cu/Fe occupation in purple. } 
\end{figure}

Successful atomistic models have been obtained by utilizing solved approximant structures and large icosahedral clusters that they contained to model hyperspace ``atomic surfaces'', then using a cut through the atomic surface to extend to true quasiperiodicity~\cite{takakura-cdyb}. Alternatively, Katz and Gratias\cite{kg93} used theoretical arguments (closeness and hard core conditions) to create atomic surfaces that maximize the density while being bounded by atomic surfaces on 2--fold planes (see Fig.~\ref{fig:perp}). Quiquandon and Gratias designed a precise chemical-ordering model for the Katz-Gratias quasilattice by atomic surface decomposition\cite{quiqua2006}.

The remainder of this section focuses on structure prediction through computer simulation. We will describe our computational techniques and present selected results together with extensive citations.

\subsection{Empirical Oscillating Pair Potentials}

Computer simulations require a model for interatomic interactions. While electronic density functional theory~\cite{HohenbergKohn,KohnSham} (DFT) provides the most accurate practical method for total energy and force calculation, it scales poorly with system size and hence is too slow to allow for simulations of large quasicrystal approximants. Machine-learning of DFT energies and forces~\cite{Behler,VASP-ML} is an emerging method that offers hope to provide near-DFT accuracy at lower computational cost. Unfortunately, the diversity of atomic environments in quasicrystals diminishes its applicability in the absence of algorithmic improvements. Instead we turn to model interatomic interaction potentials, which are functions that can be quickly evaluated to yield either the energy of a configuration or the forces acting on the atoms.

Many of the most widely studied quasicrystal-forming compounds are either Al-rich (Mackay-type), or rich in Cd or Zn (Tsai-type). Their interactions are mediated by nearly free electrons and their interaction potentials can be derived perturbatively. An ion placed into an otherwise uniform electron gas creates Friedel oscillations~\cite{Friedel1958} in the density at a frequency of $2k_F$, with $k_F$ the Fermi wavenumber. A second ion then feels an oscillating potential dependent on the local electron density. Ashcroft~\cite{AshcroftLekner} and Hafner~\cite{HafnerHam} calculated interactions for metals with $s$ and $p$ valence electrons. Moriarty~\cite{MoriartyGPT} generalized the calculations to include the $d$ electrons of transition metals and extended the theory to include many-body interactions.

\begin{figure}[tbhp]
  \centering
  \includegraphics[width=.48\textwidth]{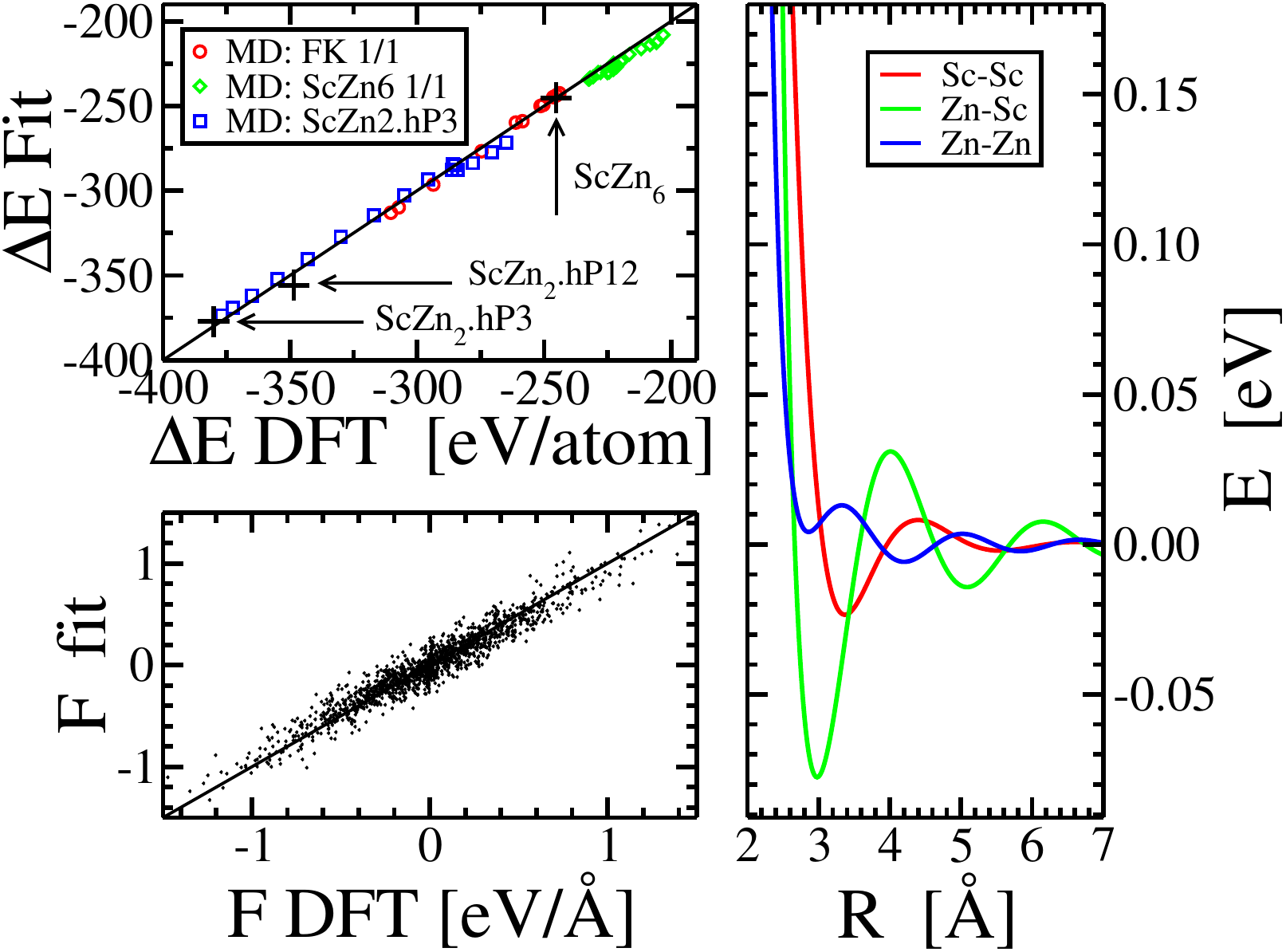}
  \caption{Fitting data and the resulting empirical oscillating pair potentials for Sc-Zn.}
  \label{fig:EOPP}
\end{figure}

Pair potentials for quasicrystal-forming compounds have been derived for Al-Co~\cite{Phillips1993}, Al-Mn~\cite{almn2}, and several other Al-TM binaries~\cite{MoriartyAlTM} and Al-Co-(Cu,Ni) ternaries~\cite{MoriartyTernary}. All of these potentials were found~\cite{EOPP} to be well-approximated by a simple functional form (Eq.~(\ref{eq:eopp})) combining Friedel-like long-range oscillations with short-range repulsion arising from overlapping ionic cores (see Fig.~\ref{fig:EOPP}),
\begin{equation}
  \label{eq:eopp}
  V(R)=\frac{C_1}{R^{\eta_1}}+\frac{C_2}{R^{\eta_2}}\cos\left(kR+\phi\right),
\end{equation}
hence the name empirical oscillating pair potential (EOPP). In practice, instead of deriving the interactions for each compound of interest, the six parameters (set independently for each chemical species pair) can be fit to a database of DFT-based energies and forces. Potentials obtained in this manner have been applied to simulations of ternary $i$-AlCuFe~\cite{i-AlCuFe}, $i$-AlMnPd~\cite{i-AlMnPd}, $i$-AlCuSc~\cite{EOPP,alcusc-ishimasa} and $d$-MgYZn~\cite{d-myz}. EOPP were also developed for many binaries. They were applied to model phonons~\cite{sczn-phonon} and low-temperature tetrahedron reshuffling\cite{sczn-tet} in $i$-ScZn and were used to resolve the vacancy distribution and identify the ground state of $\beta$-Al$_3$Mg$_2$~\cite{al3mg2}. Low temperature phase transitions in orientational order were modeled in the quasicrystal-related Al$_{11}$Ir$_4$ phase~\cite{al11ir4}.



\subsection{Simulation methods}

Given the interaction potentials, we need a method to explore the ensemble of possible structures. Because quasicrystals are aperiodic, an infinite system is required to capture the complete structure. In practice, we impose periodic boundary conditions, but the precise sizes and shapes are selected to impose only small deviations from quasiperiodicity, and to converge as quickly as possible. Taking inspiration from the optimal approximations to the golden mean provided by ratios of successive Fibonacci numbers, we follow the lead of Elser and Henley~\cite{eh85} and construct a sequence of approximants whose lattice parameters scale by powers of the golden mean, $\tau$, and label them according to Fibonacci number ratios.

For small simulation cell sizes, the special approximant cell geometries can assist in formation of quasicrystalline motifs that perfectly obey the geometrical cell constraints~\cite{CellConstraint}. For larger cell sizes, the high entropy of random liquid configurations can inhibit the solidification into the quasicrystal structure. In this case, it proves effective to seed the larger structure with a fragment of a smaller approximant that can nucleate the growth of quasicrystal throughout the larger cell~\cite{i-AlCuFe}.

\begin{figure*}[t!]
  \centering
  \includegraphics[trim=35mm 27mm 23mm 60mm, clip, width=.32\textwidth] {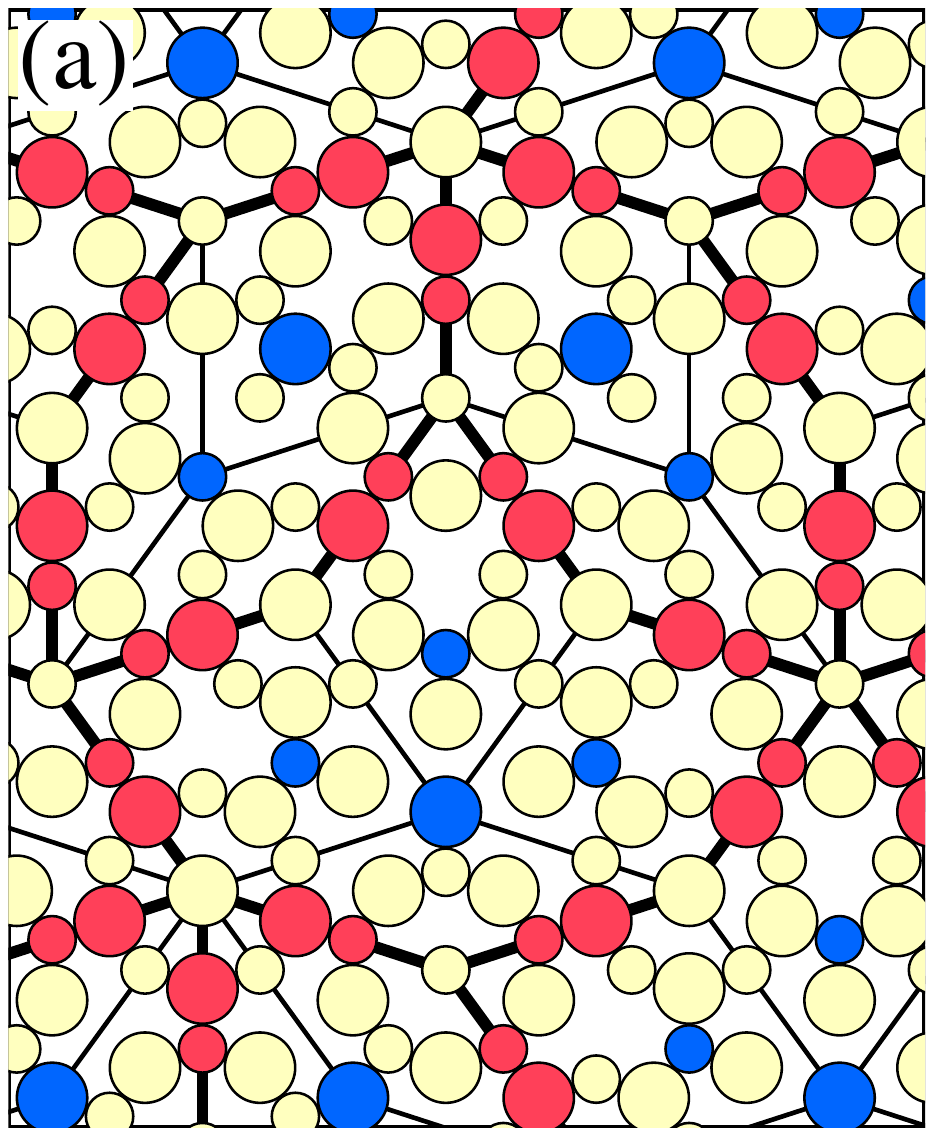}
  \includegraphics[trim=35mm 27mm 23mm 60mm, clip, width=.32\textwidth]{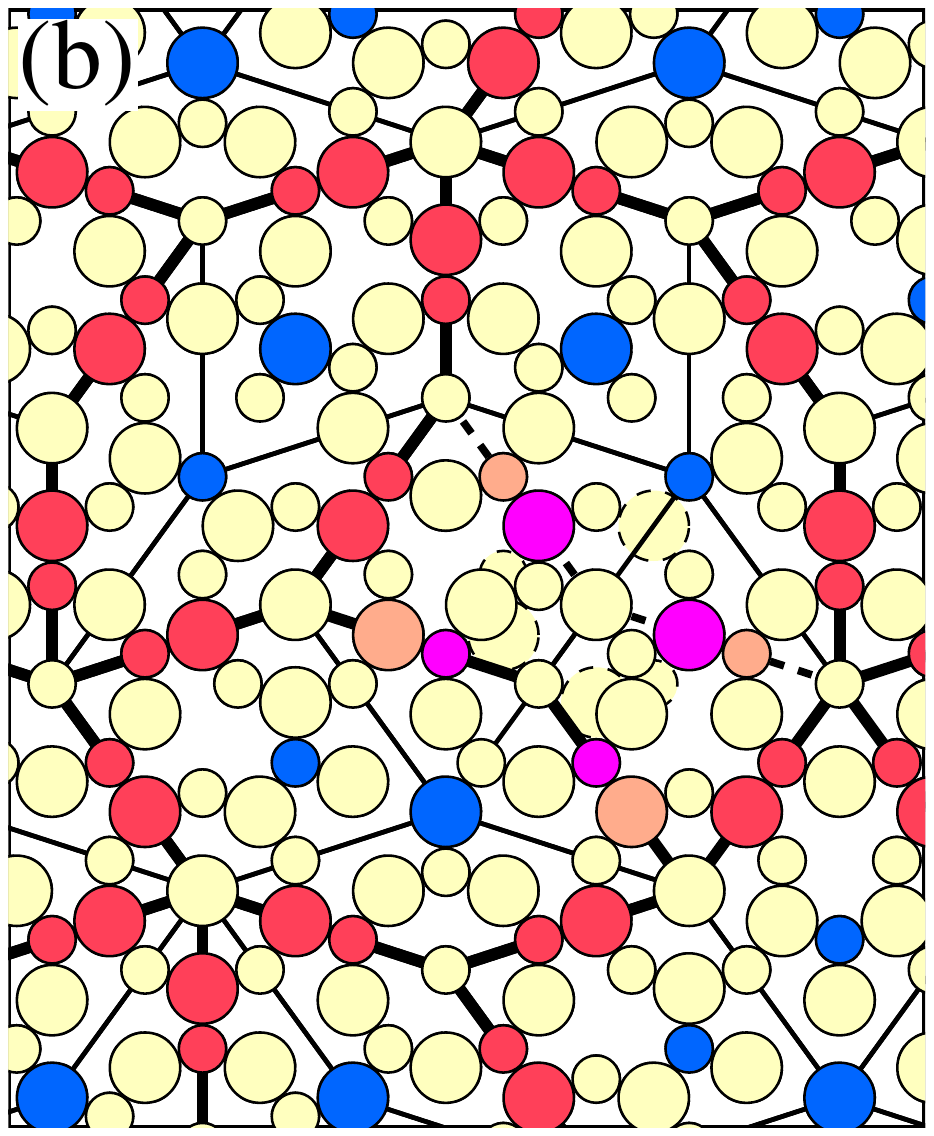}
  \includegraphics[trim=35mm 27mm 23mm 60mm, clip, width=.32\textwidth] {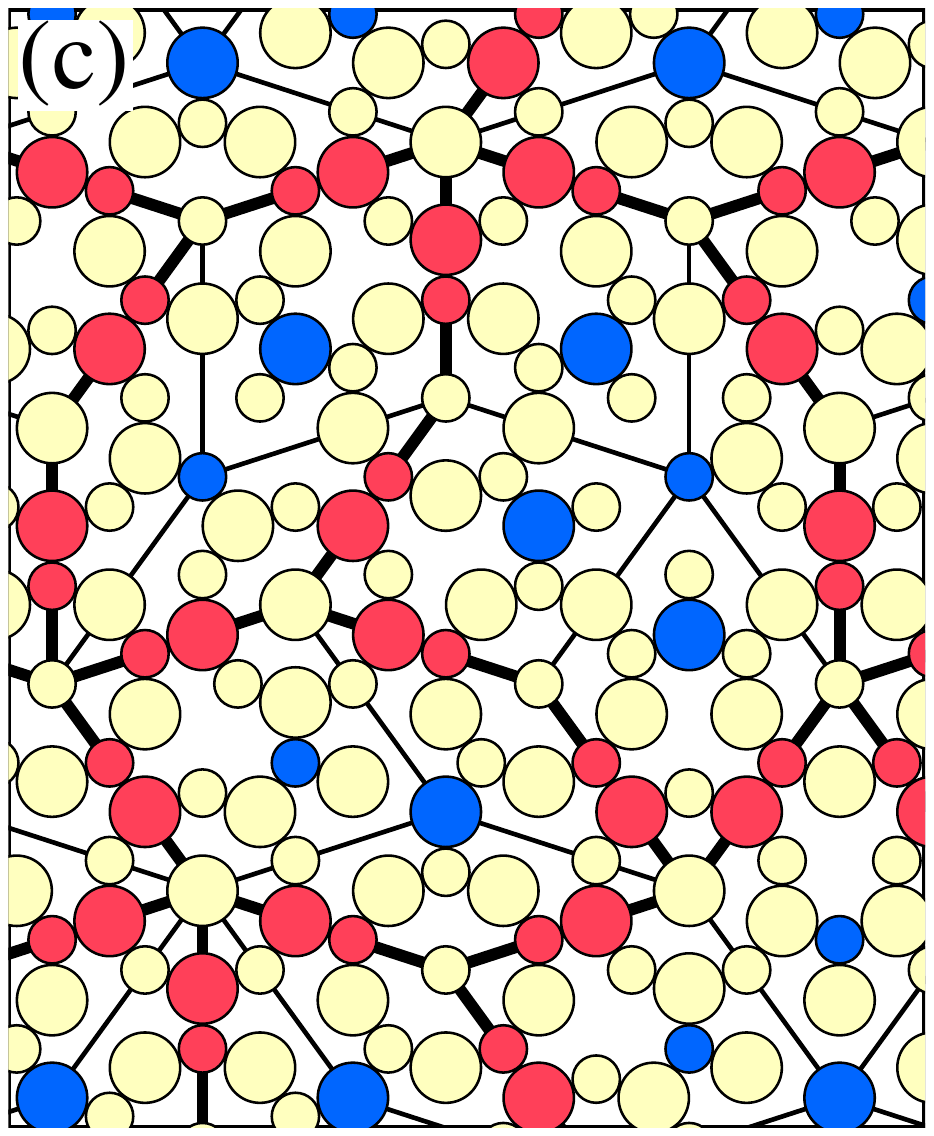}
  \caption{Species swaps and atomic displacements during phason flip in a model of $d$-AlCoNi~\cite{PhasonReview}. Tile edge length is 6.5~\AA. Atom colors are: Al (yellow), Co (blue) and Ni (red). Atom size indicates depth. Parts (a)-(c) correspond, respectively to initial HBS tiling, bowtie flip, and final tiling. In (b) orange sites correspond to swap of Al and Ni, pink corresponds to swap of Co and Ni. Dashed lines in (b) show initial tile edges and atomic positions prior to flip.}
  \label{fig:phason}
\end{figure*}

Substitutional disorder, in which pairs of pairs of atoms swap their chemical identities, is common in many quasicrystal-forming alloys. In $d$-AlCoNi, for example, Ni atoms are able to substitute for Al at many sites, and also Ni can substitute for Co. This is consistent with the location of Ni between Co and Al on the simulated atomic surface~\cite{Naidu} and in the periodic table. Unfortunately, diffusion rates in the solid state can be quite low, preventing conventional molecular dynamics from sampling the full equilibrium ensemble of structures. Thus we supplement the conventional molecular dynamics simulation with Monte Carlo steps that attempt to swap pairs of atoms of differing species. The swaps are accepted according to their Boltzmann probabilities ($\exp{(-\Delta E/\kB T)}$ with $\Delta E$ the change in energy), in order to maintain thermodynamic equilibrium. In this way a given atom can migrate over large distances without the need to cross energy barriers to diffusion.

Phasons are are a source of disorder that is unique to incommensurate structures such as quasicrystals. It is important to distinguish between phason {\em modes}, which generate long wavelength deviations from quasiperiodicity through correlated displacements of atoms~\cite{Bak1985,Lubensky1985} and localized phason {\em flips}~\cite{PhasonReview} that can be a thermodynamically stabilizing source of entropy but need not disrupt the long-range quasiperiodicity. Special Monte Carlo moves can sometimes be designed to enable barrier-free phason flips~\cite{BinaryLJ}. However, if tilings are defined by connecting atoms of certain species, then chemical species swaps followed by small relaxations are sometimes able to implement phason flips without the need to design special Monte Carlo moves. Fig.~\ref{fig:phason} illustrates a phason flip in a model of $d$-AlCoNi~\cite{PhasonReview}. The phason flip swaps the positions of a boat tile and a hexagon tile that together comprise 81 atoms. However, only four pairs of species swaps are required (two Al/Ni and two Co/Ni) along with minor relaxation of a few Al atoms. These sites of atomic species swaps are sites of mixed chemical occupancy, as indicated by their coloring in Fig.~\ref{fig:phason} part (b).

Tile-Hamiltonians~\cite{almn1,almn2,AlCoNi-deco,T-Ham-AlNiCo,T-Ham-AlCoCu-FP,T-Ham-AlCoCu} provide an efficient way to simulate quasicrystal structures based on the assumption that a quasicrystal structure is in one-to-one relationship with a tiling geometry. Energetic optimization of the tile decoration is followed by fitting a Hamiltonian that assigns energy coefficients to occurrences of preselected tiling objects. A tile-reshuffling Monte Carlo simulation then explores the tiling ensemble. For example, the tile decoration of decagonal MgZnY~\cite{d-myz} follows the well established Frank-Kasper decoration of the Rectangle-Triangle tiling~\cite{roth1997}.  The tile-Hamiltonian was fit using EOPP potentials and then applied in a Monte Carlo ``zipper''-reshuffling simulation that revealed the ground state tiling configurations in very large decagonal approximants.

We wish to carry out simulations over a range of temperatures, however equilibration is difficult to achieve at low temperature owing to the reduced swap acceptance rates. The method of replica exchange~\cite{SwendsenWang} provides a solution. Ensemble diversity is enhanced at every temperature by carrying out simulations in parallel at multiple temperatures, and swapping the configurations between temperatures using a Boltzmann-like probability $\exp{(\Delta\beta\Delta E)}$. Here $\Delta\beta$ is the difference in $1/\kB T$ and $\Delta E$ is the difference in energies of two configurations~\cite{KimWidom}. A given low temperature configuration can migrate up to higher temperatures where swapping is accelerated, then migrate back down to bring a fresh structure to the low temperature. 

Simulations can even be combined with diffraction refinement by augmenting the diffraction $R$-factor with a term dependent on the calculated total energy. This approach has been demonstrated for $d$-AlCoNi~\cite{Combined} but has not been widely adopted.

\section{Predicting thermodynamic stability of quasicrystals}

\begin{figure}[tbhp]
  \centering
  \includegraphics[width=.48\textwidth]{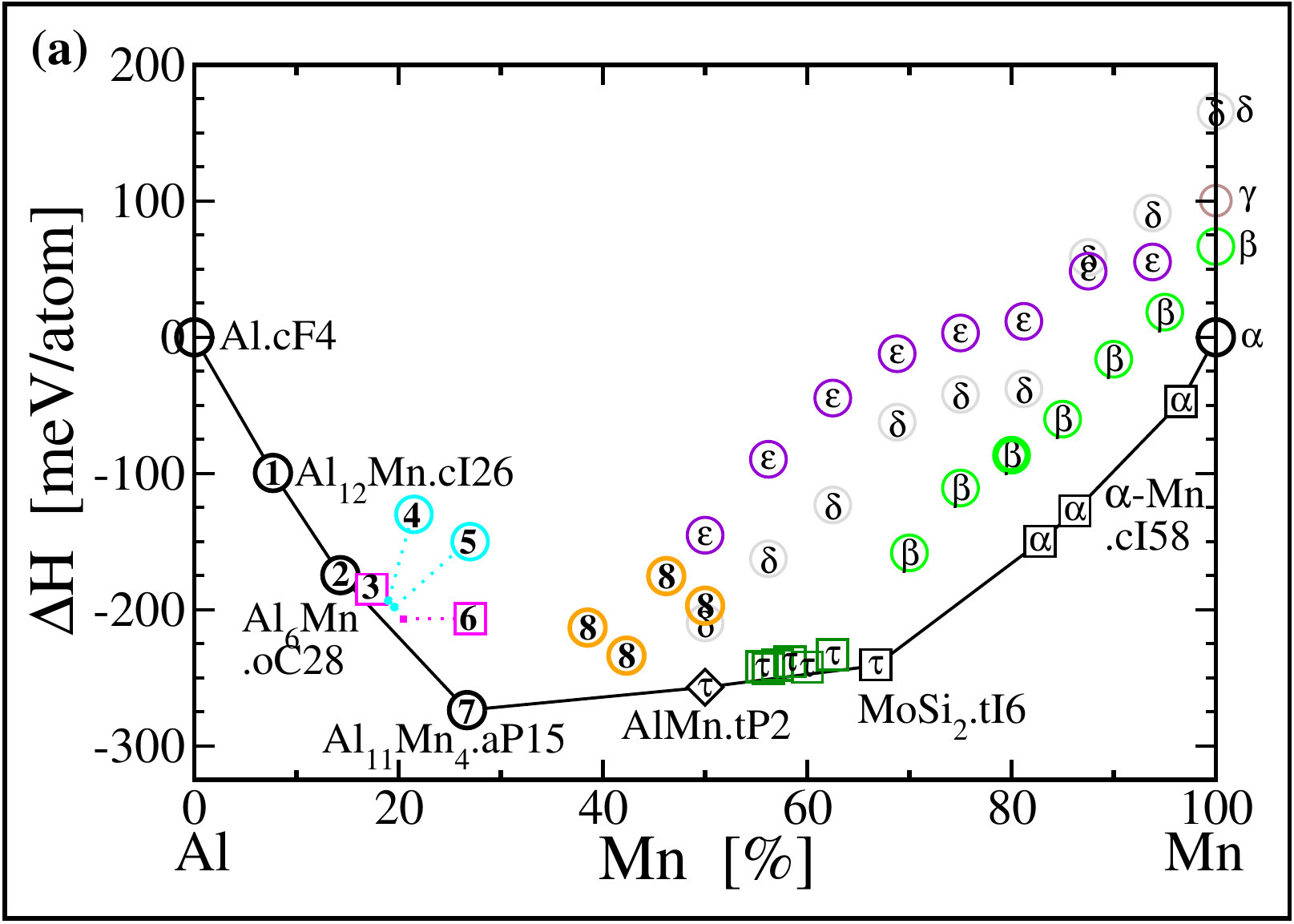}
  \includegraphics[width=.48\textwidth]{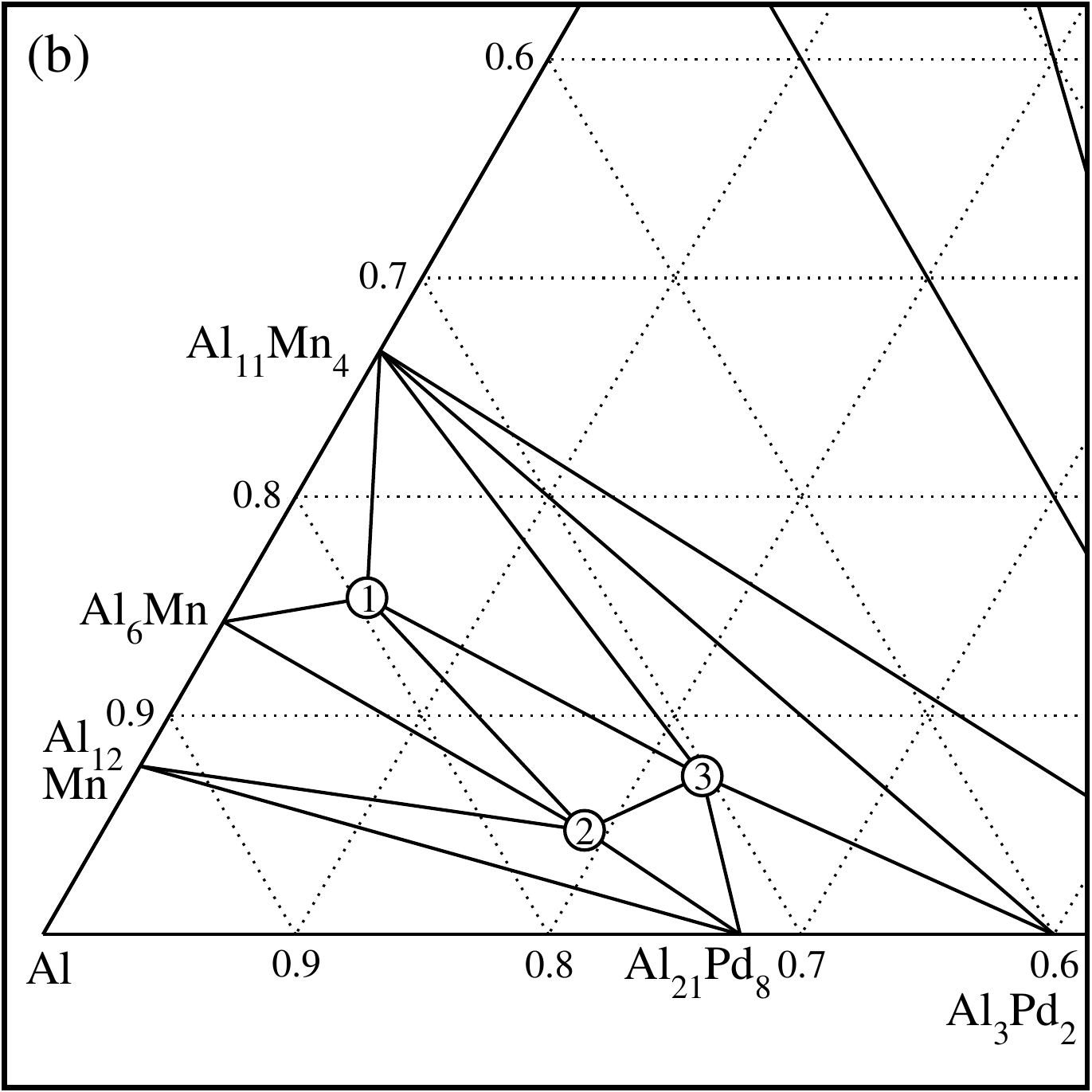}
  \caption{Convex hulls of (a) the Al-Mn binary and (b) the Al-rich region of the Al-Mn-Pd ternary~\cite{i-AlMnPd}. Stable binaries are labeled by structure type. Stable ternaries are: 12 and 16~\AA~ decagonal approximants (labeled 1 and 2) with 156 and 168 atoms, respectively, and the 3/2 icosahedral approximant (labeled 3) with 552 atoms.}
  \label{fig:Hull}
\end{figure}

Given an accurate structure model, we calculate enthalpy of formation (in the $T\to 0$K limit) by subtracting the relaxed DFT total energy from the composition-weighted average of the constituent element energies. For a structure with concentrations $\{x_\alpha\}$ for different species $\alpha$, the formation enthalpy
\begin{equation}
  \label{eq:DeltaH}
  \Delta H = E-\sum_\alpha x_\alpha E_\alpha
\end{equation}
where $E$ is the relaxed energy per atom of the given structure and $\{E_\alpha\}$ are the relaxed energies per atom of the pure elements. Consider a scatter plot of $\Delta H$ {\em vs.} composition such as Fig.~\ref{fig:Hull}a. The convex hull of this plot consists of vertices connected by facets (in this case, line segments). The vertices are predicted to be stable pure phases, while the facets identify compositions at which the adjacent pure phases coexist.

Structures whose enthalpies lie above the convex hull are predicted to be thermodynamically unstable at low temperature, however they may be mechanically stable and hence thermodynamically metastable. In some cases they may become thermodynamically stable at elevated temperatures owing to the reduction in free energy $\Delta G=\Delta H-T\Delta S$ through entropic effects. Given the ability to simulate the temperature-dependent ensemble averaged enthalpy $H(T)$, the free energy can be calculated by thermodynamic integration. First, the heat capacity may be obtained from enthalpy either by differentiation or from the averaged fluctuations,
\begin{equation}
  \label{eq:Cp}
  C_p(T) = \frac{\partial H}{\partial T}
  = \frac{1}{\kB T^2}\left(\avg{H^2} - \avg{H}^2\right).
\end{equation}
Then the entropy is obtained, up to an unknown constant $S_{\rm ref}$, by integrating,
\begin{equation}
  \label{eq:S}
  S(T) = S_{\rm ref} + \int_{T_{\rm ref}}^T \rmd T'~\frac{C_p}{T'}.
\end{equation}
Finally, we integrate once more to obtain
\begin{equation}
  \label{eq:F}
  F(T)=F_{\rm ref}-\int_{T_{\rm ref}}^T \rmd T'~ S(T')
\end{equation}
where $F_{\rm ref}=H_{\rm ref}-T S_{\rm ref}$ is an unknown linear function of $T$. A trick to determine $F_{\rm ref}$ is presented at the end of this section.

This approach can be generalized~\cite{FrenkelSmit} to yield the free energy of a parameter-dependent Hamiltonian $H(\lambda)$,
\begin{equation}
  \label{eq:FL}
  F(\lambda) = F_0 +
  \int_0^\lambda \rmd \lambda'~
  \avg{\partial H(\lambda')/\partial \lambda'},
\end{equation}
with $F_0$ the free energy for $\lambda=0$ and brackets $\avg{~}$ represent a simulated average. For example, the free energy of a hard sphere solid may be derived from a harmonic solid by this method. In the context of quasicrystals, this method was applied to calculate the phonon entropy of a Tubingen-Triangle random tiling model with Lennard-Jones-Gauss potential~\cite{Kiselev}. It was also applied to compute the phase diagram for a patchy particle system~\cite{Reinhardt} and for a hard disk system with soft corona that leads to a random square-triangle tiling~\cite{Dijkstra}. The free energy of phason fluctuations could in principle be derived by relaxing a constraint of perfect quasiperiodicity. 

\begin{figure}[tbhp]
  \centering
  \includegraphics[width=.48\textwidth]{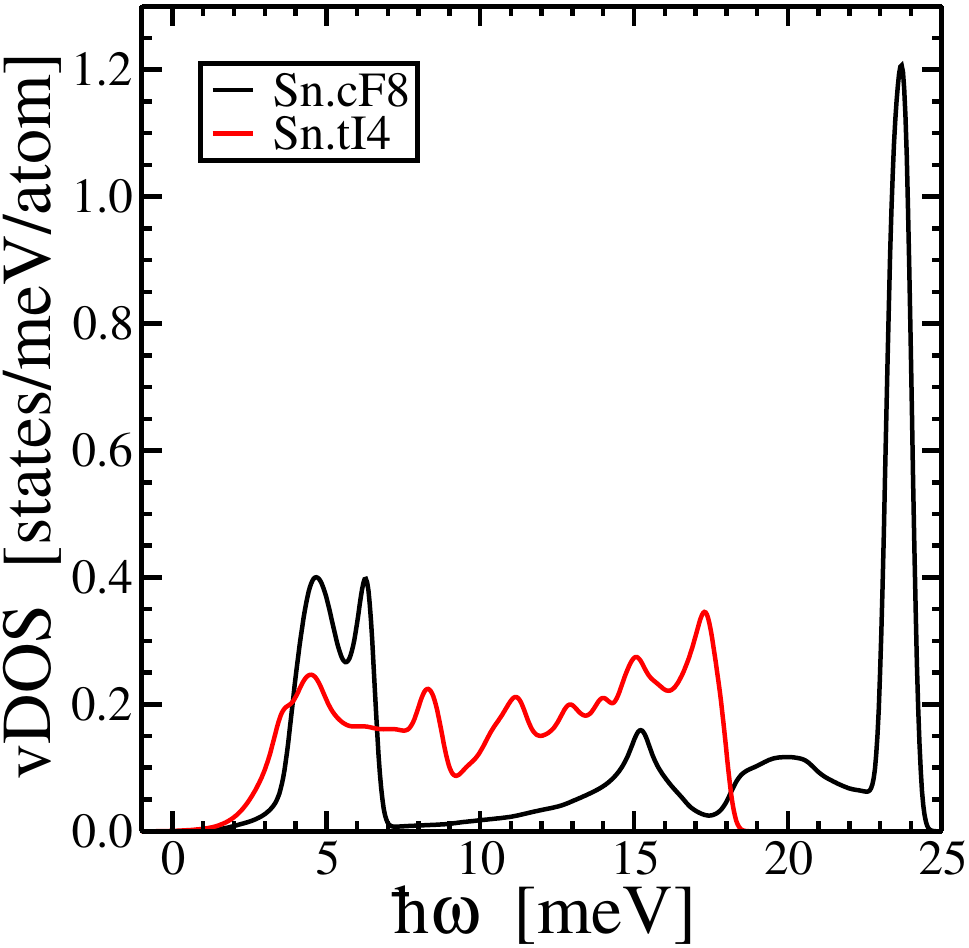}
  \caption{Vibrational densities of states of diamond ($\alpha$-Sn.cF8) and bct ($\beta$-Sn.tI2) tin.}
  \label{fig:Fvib}
\end{figure}

Many entropy sources contribute to the free energy. The largest in magnitude is usually the vibrational entropy~\cite{Fultz2010}, and it is often capable of stabilizing structures that would be metastable~\cite{Wolverton2001} or even mechanically unstable~\cite{Svib} at low temperatures. In the harmonic approximation the vibrational free energy is
\begin{equation}
  \label{eq:Fvib}
  F_{\rm vib} = \kB T \int{\rm d}\omega~ g(\omega) \ln{[2\sinh(\hbar\omega/2k_{\rm B}T)]}.
\end{equation}
Low frequency modes with $\hbar\omega\ll \kB T$ reduce the vibrational free energy by approximately $\kB T\ln{(\hbar\omega/\kB T)}$, while high frequency modes {\em increase} the quantum zero point energy by $\hbar\omega/2$. As an example of stabilization by vibrational entropy, elemental tin (Sn) transforms from the diamond structure ($\alpha$-Sn.cF8) to body-centered tetragonal ($\beta$-Sn.tI4) at 13$^\circ$C (286$^\circ$K, $\kB T=25$ meV). As shown in Fig.~\ref{fig:Fvib}, all the vibrational states of $\beta$ are excited by this temperature, and its relatively low frequencies stabilize $\beta$ over $\alpha$.

Excitations of electrons from occupied states below the Fermi energy to empty states above also contribute to reducing the free energy. The free energy reduction is approximately~\cite{Chapter8}
\begin{equation}
  \label{eq:Felect}
  F_e(T) \approx -\frac{\pi^2}{6} D(E_F) (\kB T)^2.
\end{equation}
Because of the quadratic dependence on $\kB T$, this effect is typically quite small at low or moderate temperatures. For example, a $D(E_F)$ of one electronic state/eV/atom reduces the free energy by only one meV/atom at room temperature. Like many intermetallics, quasicrystals often have deep pseudogaps, making the electronic contribution negligible except at elevated temperatures. In the case of $i$-AlCuFe, $D(E_F)\approx0.2$ states/eV/atom, resulting in $F_e\approx -0.2$ meV/atom at room temperature, and $-3.5$ meV/atom at 1200K.

The unique entropic contribution to quasicrystals comes from phason fluctuations (see Fig.~\ref{fig:phason}). The precise identification of a phason in a realistic quasicrystal structure is ill-defined; they are frequently defined in terms of tile flips, but tilings can be identified on many different length scales. For example, the idealized $d$-AlCoNi structure in Fig.~\ref{fig:phason} is derived from an HBS tiling with a smaller 3~\AA~ edge length, that can in turn be decomposed into a rhombus tiling with 2.45~\AA~ edges~\cite{AlCoNi-deco}. The tilings themselves are often defined in terms of specific chemical species, but species swaps are prevalent. Phasons even create small atomic displacements and hence are linked with diffusion.

Perhaps the clearest manifestation of phasons is through the atomic surface occupation densities $\{\rho_\alpha(\br_\perp)\}$ that show mixed and partial occupation of species $\alpha$ as functions of perpendicular space position $\br_\perp$. In $i$-AlCuFe (Fig.~\ref{fig:perp}), swapping an individual Fe atom with Cu or Cu with Al corresponds to a perpendicular space shift of the atomic surface centers, justifying the link between species swaps and localized phason flips. In principle the entropy could be calculated from the densities as
\begin{equation}
  \label{eq:S-info}
  S_{\rm phason} = - \sum_\alpha \int_{\rm AS}\rmd~\br_\perp~
 \rho_\alpha(\br_\perp) \ln{\rho_\alpha(\br_\perp)}
\end{equation}
where the densities are normalized so that $\sum_\alpha\rho_\alpha(\br_\perp)$ reaches a maximum value of 1. This method has not yet been applied, although a functional representation of $\rho_\alpha(\br_\perp)$ is available~\cite{Naidu}

\begin{figure}[tbhp]
  \centering
  \includegraphics[width=.48\textwidth]{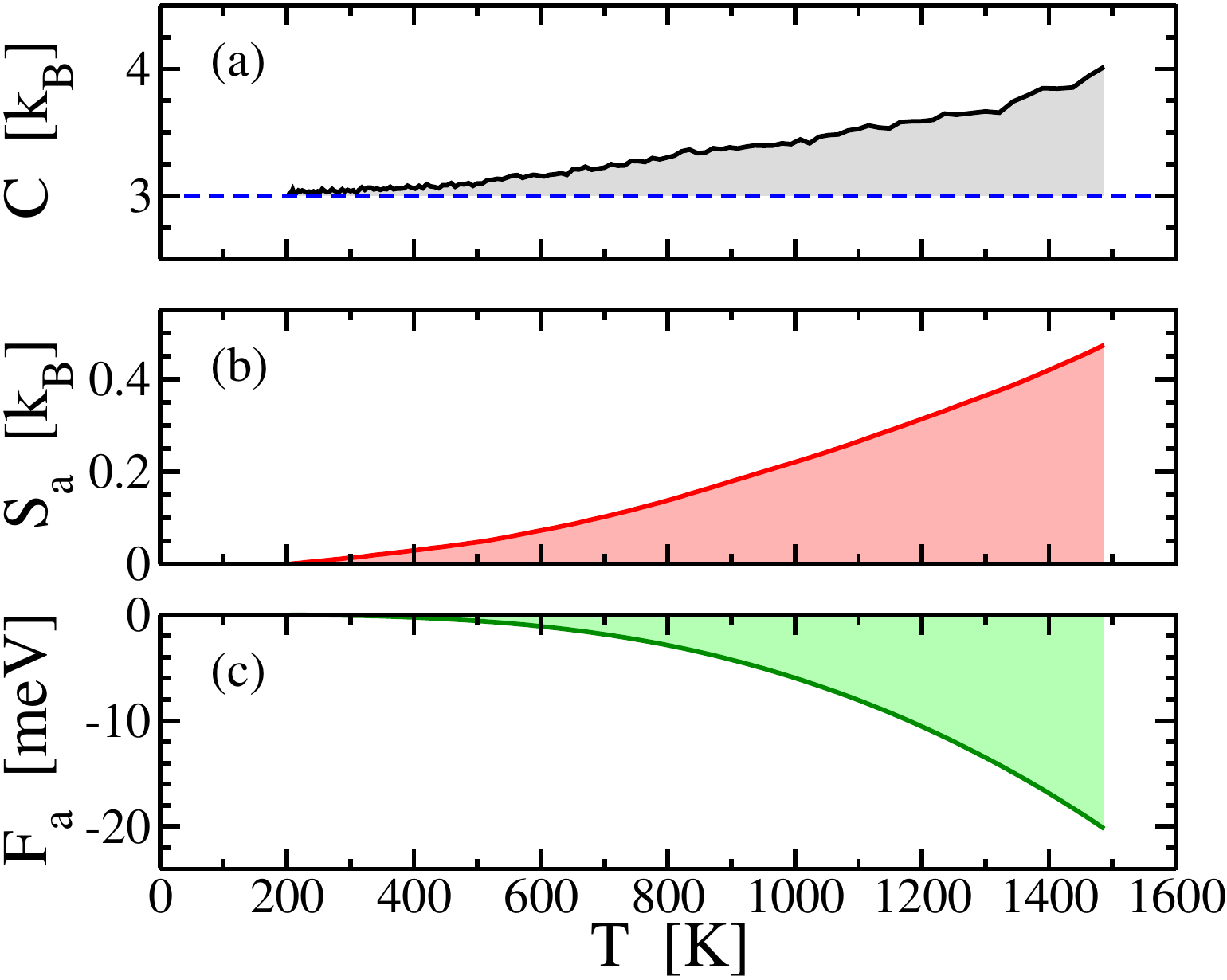}
  \caption{\label{fig:Ca} Anharmonic thermodynamics of $i$-AlCuFe. (a) Heat capacity from MC/MD simulation; shaded area is anharmonic part, $C_a$, in units of $\kB$/atom. (b) Anharmonic entropy in $\kB$/atom. (c) Anharmonic free energy in meV/atom. }
\end{figure}

In our study of $i$-AlCuFe we took a different approach based separating harmonic and anharmonic contributions to the free energy. We defined an ``anharmonic heat capacity'', $C_a(T)=C(T)-3\kB$, where $C(T)$ is the heat capacity measured during our MC/MD simulation and $3\kB$ is the classical harmonic value of Dulong and Petit. Thermodynamic integration of the anharmonic heat capacity via Eq.~(\ref{eq:S}) yields an ``anharmonic entropy'', $S_a(T)$, and further integration via Eq.~(\ref{eq:F}) yields an ``anharmonic free energy'', $F_a(T)$. As shown in Figure~\ref{fig:Ca}a, $C_a$ vanishes faster than linearly at low temperatures. Assuming that $S_a$ also vanishes at low temperature, we may set $F_{a, \rm ref}=0$. The resulting thermodynamic functions shown in Fig.~\ref{fig:Ca}b and c can be directly added to the corresponding harmonic phonon functions. Notice that the anharmonic contribution reduces the free energy by up to 20 meV/atom at high temperature, which is substantially larger than the $\Delta E$ values of typical quasicrystal approximants relative to competing phases, although it is an order of magnitude smaller than the harmonic vibrational free energy.

\section{Summary and outlook}

In the years since their first discovery, tremendous strides have been made in the prediction of quasicrystal structures. While true {\em ab-initio} prediction of atomistic quasicrystal formation in a given alloy system has fallen short, new emerging frontiers creating artificial quasicrystals might greatly expand the array of systems with quasicrystalline structures. Promising avenues include the design of interparticle interactions to encourage quasiperiodicity or non-crystallographic rotational symmetry in colloids and other soft-matter systems. Alternatively quasiperiodic potentials can be created through superpositions of waves, and electrons, atoms, or colloidal particles follow the imposed potential.

For known quasicrystal-forming metallic alloys, the problem of predicting their structure is closely tied with the problem of understanding their formation because the structures are governed by the same chemical interactions that bind the elements into a solid. At the same time, detailed structure models are required in order to calculate accurate energies. Both tasks are aided by the growing capabilities of total energy calculation and computer simulation methods. Today, chemically accurate interaction energies can be calculated directly from first principles for quasicrystal approximants with thousands of atoms. Simulations that explore configurational ensembles require greater speed. Existing interatomic potential models are often sufficiently accurate for simulation of high temperature properties, and it is hoped that machine learning approaches in the future could achieve even more accurate potentials, enabling ground state searches.

\section{Acknowledgements}
MM is thankful for the support from the Slovak Grant Agency  VEGA (No. 2/0144/21) and APVV (No. 20-0124, No. 19-0369). MW was supported by the Department of Energy under Grant No. DE-SC0014506.

\bibliography{refs}

\begin{thebibliography}{81}%
\makeatletter
\providecommand \@ifxundefined [1]{%
 \@ifx{#1\undefined}
}%
\providecommand \@ifnum [1]{%
 \ifnum #1\expandafter \@firstoftwo
 \else \expandafter \@secondoftwo
 \fi
}%
\providecommand \@ifx [1]{%
 \ifx #1\expandafter \@firstoftwo
 \else \expandafter \@secondoftwo
 \fi
}%
\providecommand \natexlab [1]{#1}%
\providecommand \enquote  [1]{``#1''}%
\providecommand \bibnamefont  [1]{#1}%
\providecommand \bibfnamefont [1]{#1}%
\providecommand \citenamefont [1]{#1}%
\providecommand \href@noop [0]{\@secondoftwo}%
\providecommand \href [0]{\begingroup \@sanitize@url \@href}%
\providecommand \@href[1]{\@@startlink{#1}\@@href}%
\providecommand \@@href[1]{\endgroup#1\@@endlink}%
\providecommand \@sanitize@url [0]{\catcode `\\12\catcode `\$12\catcode
  `\&12\catcode `\#12\catcode `\^12\catcode `\_12\catcode `\%12\relax}%
\providecommand \@@startlink[1]{}%
\providecommand \@@endlink[0]{}%
\providecommand \url  [0]{\begingroup\@sanitize@url \@url }%
\providecommand \@url [1]{\endgroup\@href {#1}{\urlprefix }}%
\providecommand \urlprefix  [0]{URL }%
\providecommand \Eprint [0]{\href }%
\providecommand \doibase [0]{http://dx.doi.org/}%
\providecommand \selectlanguage [0]{\@gobble}%
\providecommand \bibinfo  [0]{\@secondoftwo}%
\providecommand \bibfield  [0]{\@secondoftwo}%
\providecommand \translation [1]{[#1]}%
\providecommand \BibitemOpen [0]{}%
\providecommand \bibitemStop [0]{}%
\providecommand \bibitemNoStop [0]{.\EOS\space}%
\providecommand \EOS [0]{\spacefactor3000\relax}%
\providecommand \BibitemShut  [1]{\csname bibitem#1\endcsname}%
\let\auto@bib@innerbib\@empty
\bibitem [{\citenamefont {Penrose}(1974)}]{Penrose1974}%
  \BibitemOpen
  \bibfield  {author} {\bibinfo {author} {\bibfnamefont {R.}~\bibnamefont
  {Penrose}},\ }\href@noop {} {\bibfield  {journal} {\bibinfo  {journal} {Bul.
  Inst. Math. Appl.}\ }\textbf {\bibinfo {volume} {10}},\ \bibinfo {pages}
  {266} (\bibinfo {year} {1974})}\BibitemShut {NoStop}%
\bibitem [{\citenamefont {Gardner}(1977)}]{Gardner1977}%
  \BibitemOpen
  \bibfield  {author} {\bibinfo {author} {\bibfnamefont {M.}~\bibnamefont
  {Gardner}},\ }\href@noop {} {\bibfield  {journal} {\bibinfo  {journal} {Sci.
  Am.}\ }\textbf {\bibinfo {volume} {236}},\ \bibinfo {pages} {110} (\bibinfo
  {year} {1977})}\BibitemShut {NoStop}%
\bibitem [{\citenamefont {Mackay}(1982)}]{Mackay1982}%
  \BibitemOpen
  \bibfield  {author} {\bibinfo {author} {\bibfnamefont {A.~L.}\ \bibnamefont
  {Mackay}},\ }\href {\doibase https://doi.org/10.1016/0378-4371(82)90359-4}
  {\bibfield  {journal} {\bibinfo  {journal} {Physica A: Statistical Mechanics
  and its Applications}\ }\textbf {\bibinfo {volume} {114}},\ \bibinfo {pages}
  {609} (\bibinfo {year} {1982})}\BibitemShut {NoStop}%
\bibitem [{\citenamefont {Shechtman}\ \emph {et~al.}(1984)\citenamefont
  {Shechtman}, \citenamefont {Blech}, \citenamefont {Gratias},\ and\
  \citenamefont {Cahn}}]{Shechtman1984}%
  \BibitemOpen
  \bibfield  {author} {\bibinfo {author} {\bibfnamefont {D.}~\bibnamefont
  {Shechtman}}, \bibinfo {author} {\bibfnamefont {I.}~\bibnamefont {Blech}},
  \bibinfo {author} {\bibfnamefont {D.}~\bibnamefont {Gratias}}, \ and\
  \bibinfo {author} {\bibfnamefont {J.~W.}\ \bibnamefont {Cahn}},\ }\href
  {\doibase 10.1103/PhysRevLett.53.1951} {\bibfield  {journal} {\bibinfo
  {journal} {Phys. Rev. Lett.}\ }\textbf {\bibinfo {volume} {53}},\ \bibinfo
  {pages} {1951} (\bibinfo {year} {1984})}\BibitemShut {NoStop}%
\bibitem [{\citenamefont {Levine}\ and\ \citenamefont
  {Steinhardt}(1984)}]{Levine1984}%
  \BibitemOpen
  \bibfield  {author} {\bibinfo {author} {\bibfnamefont {D.}~\bibnamefont
  {Levine}}\ and\ \bibinfo {author} {\bibfnamefont {P.~J.}\ \bibnamefont
  {Steinhardt}},\ }\href@noop {} {\bibfield  {journal} {\bibinfo  {journal}
  {Phys. Rev. Lett.}\ }\textbf {\bibinfo {volume} {53}},\ \bibinfo {pages}
  {2477} (\bibinfo {year} {1984})}\BibitemShut {NoStop}%
\bibitem [{\citenamefont {Dunlap}\ and\ \citenamefont
  {Dini}(1985)}]{Dunlap1985}%
  \BibitemOpen
  \bibfield  {author} {\bibinfo {author} {\bibfnamefont {R.~A.}\ \bibnamefont
  {Dunlap}}\ and\ \bibinfo {author} {\bibfnamefont {K.}~\bibnamefont {Dini}},\
  }\href@noop {} {\bibfield  {journal} {\bibinfo  {journal} {Can. J. Phys.}\
  }\textbf {\bibinfo {volume} {63}},\ \bibinfo {pages} {1267} (\bibinfo {year}
  {1985})}\BibitemShut {NoStop}%
\bibitem [{\citenamefont {Dubois}\ \emph {et~al.}(1986)\citenamefont {Dubois},
  \citenamefont {Janot},\ and\ \citenamefont {Pannetier}}]{Dubois1986}%
  \BibitemOpen
  \bibfield  {author} {\bibinfo {author} {\bibfnamefont {J.~M.}\ \bibnamefont
  {Dubois}}, \bibinfo {author} {\bibfnamefont {C.}~\bibnamefont {Janot}}, \
  and\ \bibinfo {author} {\bibfnamefont {J.}~\bibnamefont {Pannetier}},\
  }\href@noop {} {\bibfield  {journal} {\bibinfo  {journal} {Phys. Lett. A}\
  }\textbf {\bibinfo {volume} {115}},\ \bibinfo {pages} {177} (\bibinfo {year}
  {1986})}\BibitemShut {NoStop}%
\bibitem [{\citenamefont {Widom}(2017)}]{Widom2017}%
  \BibitemOpen
  \bibfield  {author} {\bibinfo {author} {\bibfnamefont {M.}~\bibnamefont
  {Widom}},\ }\href@noop {} {\bibfield  {journal} {\bibinfo  {journal} {J.
  Stat. Phys.}\ }\textbf {\bibinfo {volume} {167}},\ \bibinfo {pages} {726}
  (\bibinfo {year} {2017})}\BibitemShut {NoStop}%
\bibitem [{\citenamefont {Bradley}\ and\ \citenamefont
  {Goldschmidt}(1939)}]{Bradley1939}%
  \BibitemOpen
  \bibfield  {author} {\bibinfo {author} {\bibfnamefont {A.~J.}\ \bibnamefont
  {Bradley}}\ and\ \bibinfo {author} {\bibfnamefont {H.}~\bibnamefont
  {Goldschmidt}},\ }\href@noop {} {\bibfield  {journal} {\bibinfo  {journal}
  {J. Inst. Met.}\ }\textbf {\bibinfo {volume} {65}},\ \bibinfo {pages} {403}
  (\bibinfo {year} {1939})}\BibitemShut {NoStop}%
\bibitem [{\citenamefont {Hardy}\ and\ \citenamefont
  {Silcox}(1955)}]{hardy1955}%
  \BibitemOpen
  \bibfield  {author} {\bibinfo {author} {\bibfnamefont {H.~K.}\ \bibnamefont
  {Hardy}}\ and\ \bibinfo {author} {\bibfnamefont {J.~M.}\ \bibnamefont
  {Silcox}},\ }\href@noop {} {\bibfield  {journal} {\bibinfo  {journal} {J.
  Inst. Metall.}\ }\textbf {\bibinfo {volume} {24}},\ \bibinfo {pages} {423}
  (\bibinfo {year} {1955})}\BibitemShut {NoStop}%
\bibitem [{\citenamefont {Dubost}\ \emph {et~al.}(1986)\citenamefont {Dubost},
  \citenamefont {Lang}, \citenamefont {Tanaka}, \citenamefont {Sainfort},\ and\
  \citenamefont {Audier}}]{Dubost1986}%
  \BibitemOpen
  \bibfield  {author} {\bibinfo {author} {\bibfnamefont {B.}~\bibnamefont
  {Dubost}}, \bibinfo {author} {\bibfnamefont {J.-M.}\ \bibnamefont {Lang}},
  \bibinfo {author} {\bibfnamefont {M.}~\bibnamefont {Tanaka}}, \bibinfo
  {author} {\bibfnamefont {P.}~\bibnamefont {Sainfort}}, \ and\ \bibinfo
  {author} {\bibfnamefont {M.}~\bibnamefont {Audier}},\ }\href@noop {}
  {\bibfield  {journal} {\bibinfo  {journal} {Nature}\ }\textbf {\bibinfo
  {volume} {324}},\ \bibinfo {pages} {48} (\bibinfo {year} {1986})}\BibitemShut
  {NoStop}%
\bibitem [{\citenamefont {Palenzona}(1971)}]{palenzona71}%
  \BibitemOpen
  \bibfield  {author} {\bibinfo {author} {\bibfnamefont {A.}~\bibnamefont
  {Palenzona}},\ }\href {\doibase https://doi.org/10.1016/0022-5088(71)90179-2}
  {\bibfield  {journal} {\bibinfo  {journal} {J. Less Common Met.}\ }\textbf
  {\bibinfo {volume} {25}},\ \bibinfo {pages} {367} (\bibinfo {year}
  {1971})}\BibitemShut {NoStop}%
\bibitem [{\citenamefont {Tsai}\ \emph {et~al.}(2000)\citenamefont {Tsai},
  \citenamefont {Guo}, \citenamefont {Abe}, \citenamefont {Takakura},\ and\
  \citenamefont {Sato}}]{tsai2000}%
  \BibitemOpen
  \bibfield  {author} {\bibinfo {author} {\bibfnamefont {A.-P.}\ \bibnamefont
  {Tsai}}, \bibinfo {author} {\bibfnamefont {J.}~\bibnamefont {Guo}}, \bibinfo
  {author} {\bibfnamefont {E.}~\bibnamefont {Abe}}, \bibinfo {author}
  {\bibfnamefont {H.}~\bibnamefont {Takakura}}, \ and\ \bibinfo {author}
  {\bibfnamefont {T.~J.}\ \bibnamefont {Sato}},\ }\href@noop {} {\bibfield
  {journal} {\bibinfo  {journal} {Nature}\ }\textbf {\bibinfo {volume} {408}},\
  \bibinfo {pages} {537} (\bibinfo {year} {2000})}\BibitemShut {NoStop}%
\bibitem [{\citenamefont {Elser}\ and\ \citenamefont {Henley}(1985)}]{eh85}%
  \BibitemOpen
  \bibfield  {author} {\bibinfo {author} {\bibfnamefont {V.}~\bibnamefont
  {Elser}}\ and\ \bibinfo {author} {\bibfnamefont {C.~L.}\ \bibnamefont
  {Henley}},\ }\href@noop {} {\bibfield  {journal} {\bibinfo  {journal} {Phys.
  Rev. Lett.}\ }\textbf {\bibinfo {volume} {55}},\ \bibinfo {pages} {2883}
  (\bibinfo {year} {1985})}\BibitemShut {NoStop}%
\bibitem [{\citenamefont {Takeda}\ \emph {et~al.}(1993)\citenamefont {Takeda},
  \citenamefont {Kimura}, \citenamefont {Hori}, \citenamefont {Yamashita},\
  and\ \citenamefont {Ino}}]{KimuraB}%
  \BibitemOpen
  \bibfield  {author} {\bibinfo {author} {\bibfnamefont {M.}~\bibnamefont
  {Takeda}}, \bibinfo {author} {\bibfnamefont {K.}~\bibnamefont {Kimura}},
  \bibinfo {author} {\bibfnamefont {A.}~\bibnamefont {Hori}}, \bibinfo {author}
  {\bibfnamefont {H.}~\bibnamefont {Yamashita}}, \ and\ \bibinfo {author}
  {\bibfnamefont {H.}~\bibnamefont {Ino}},\ }\href@noop {} {\bibfield
  {journal} {\bibinfo  {journal} {Phys. Rev. B}\ }\textbf {\bibinfo {volume}
  {48}},\ \bibinfo {pages} {13159} (\bibinfo {year} {1993})}\BibitemShut
  {NoStop}%
\bibitem [{\citenamefont {Ramachandrarao}\ and\ \citenamefont
  {Sastry}(1985)}]{FK}%
  \BibitemOpen
  \bibfield  {author} {\bibinfo {author} {\bibfnamefont {P.}~\bibnamefont
  {Ramachandrarao}}\ and\ \bibinfo {author} {\bibfnamefont {G.}~\bibnamefont
  {Sastry}},\ }\href@noop {} {\bibfield  {journal} {\bibinfo  {journal}
  {Pramana}\ }\textbf {\bibinfo {volume} {25}},\ \bibinfo {pages} {L225}
  (\bibinfo {year} {1985})}\BibitemShut {NoStop}%
\bibitem [{\citenamefont {Liu}\ \emph {et~al.}(2021)\citenamefont {Liu},
  \citenamefont {Fujita}, \citenamefont {Katsura}, \citenamefont {Inada},
  \citenamefont {Ishikawa}, \citenamefont {Tamura}, \citenamefont {Kimura},\
  and\ \citenamefont {Yoshida}}]{KimuraML}%
  \BibitemOpen
  \bibfield  {author} {\bibinfo {author} {\bibfnamefont {C.}~\bibnamefont
  {Liu}}, \bibinfo {author} {\bibfnamefont {E.}~\bibnamefont {Fujita}},
  \bibinfo {author} {\bibfnamefont {Y.}~\bibnamefont {Katsura}}, \bibinfo
  {author} {\bibfnamefont {Y.}~\bibnamefont {Inada}}, \bibinfo {author}
  {\bibfnamefont {A.}~\bibnamefont {Ishikawa}}, \bibinfo {author}
  {\bibfnamefont {R.}~\bibnamefont {Tamura}}, \bibinfo {author} {\bibfnamefont
  {K.}~\bibnamefont {Kimura}}, \ and\ \bibinfo {author} {\bibfnamefont
  {R.}~\bibnamefont {Yoshida}},\ }\href@noop {} {\bibfield  {journal} {\bibinfo
   {journal} {Adv. Mat.}\ }\textbf {\bibinfo {volume} {33}},\ \bibinfo {pages}
  {2102507} (\bibinfo {year} {2021})}\BibitemShut {NoStop}%
\bibitem [{\citenamefont {Miyazaki}\ \emph {et~al.}(2010)\citenamefont
  {Miyazaki}, \citenamefont {Okada}, \citenamefont {Abe}, \citenamefont
  {Yokoyama},\ and\ \citenamefont {Kimura}}]{KimuraBTiRu}%
  \BibitemOpen
  \bibfield  {author} {\bibinfo {author} {\bibfnamefont {Y.}~\bibnamefont
  {Miyazaki}}, \bibinfo {author} {\bibfnamefont {J.}~\bibnamefont {Okada}},
  \bibinfo {author} {\bibfnamefont {E.}~\bibnamefont {Abe}}, \bibinfo {author}
  {\bibfnamefont {Y.}~\bibnamefont {Yokoyama}}, \ and\ \bibinfo {author}
  {\bibfnamefont {K.}~\bibnamefont {Kimura}},\ }\href@noop {} {\bibfield
  {journal} {\bibinfo  {journal} {J. Phys. Soc. Jpn.}\ }\textbf {\bibinfo
  {volume} {79}},\ \bibinfo {pages} {073601} (\bibinfo {year}
  {2010})}\BibitemShut {NoStop}%
\bibitem [{\citenamefont {Mihalkovi\ifmmode~\check{c}\else \v{c}\fi{}}\ and\
  \citenamefont {Widom}(2004)}]{BMgRu}%
  \BibitemOpen
  \bibfield  {author} {\bibinfo {author} {\bibfnamefont {M.}~\bibnamefont
  {Mihalkovi\ifmmode~\check{c}\else \v{c}\fi{}}}\ and\ \bibinfo {author}
  {\bibfnamefont {M.}~\bibnamefont {Widom}},\ }\href {\doibase
  10.1103/PhysRevLett.93.095507} {\bibfield  {journal} {\bibinfo  {journal}
  {Phys. Rev. Lett.}\ }\textbf {\bibinfo {volume} {93}},\ \bibinfo {pages}
  {095507} (\bibinfo {year} {2004})}\BibitemShut {NoStop}%
\bibitem [{\citenamefont {Lancon}\ \emph {et~al.}(1986)\citenamefont {Lancon},
  \citenamefont {Billard},\ and\ \citenamefont {Chaudhari}}]{Lancon}%
  \BibitemOpen
  \bibfield  {author} {\bibinfo {author} {\bibfnamefont {F.}~\bibnamefont
  {Lancon}}, \bibinfo {author} {\bibfnamefont {L.}~\bibnamefont {Billard}}, \
  and\ \bibinfo {author} {\bibfnamefont {P.}~\bibnamefont {Chaudhari}},\ }\href
  {\doibase 10.1209/0295-5075/2/8/009} {\bibfield  {journal} {\bibinfo
  {journal} {Europhys. Lett.}\ }\textbf {\bibinfo {volume} {2}},\ \bibinfo
  {pages} {625} (\bibinfo {year} {1986})}\BibitemShut {NoStop}%
\bibitem [{\citenamefont {Widom}\ \emph {et~al.}(1987)\citenamefont {Widom},
  \citenamefont {Strandburg},\ and\ \citenamefont {Swendsen}}]{BinaryLJ}%
  \BibitemOpen
  \bibfield  {author} {\bibinfo {author} {\bibfnamefont {M.}~\bibnamefont
  {Widom}}, \bibinfo {author} {\bibfnamefont {K.~J.}\ \bibnamefont
  {Strandburg}}, \ and\ \bibinfo {author} {\bibfnamefont {R.~H.}\ \bibnamefont
  {Swendsen}},\ }\href@noop {} {\bibfield  {journal} {\bibinfo  {journal}
  {Phys. Rev. Lett.}\ }\textbf {\bibinfo {volume} {58}},\ \bibinfo {pages}
  {706} (\bibinfo {year} {1987})}\BibitemShut {NoStop}%
\bibitem [{\citenamefont {Dzugutov}(1992)}]{Dzugutov}%
  \BibitemOpen
  \bibfield  {author} {\bibinfo {author} {\bibfnamefont {M.}~\bibnamefont
  {Dzugutov}},\ }\href {\doibase 10.1103/PhysRevA.46.R2984} {\bibfield
  {journal} {\bibinfo  {journal} {Phys. Rev. A}\ }\textbf {\bibinfo {volume}
  {46}},\ \bibinfo {pages} {R2984} (\bibinfo {year} {1992})}\BibitemShut
  {NoStop}%
\bibitem [{\citenamefont {Kiselev}\ \emph {et~al.}(2012)\citenamefont
  {Kiselev}, \citenamefont {Engel},\ and\ \citenamefont {Trebin}}]{Kiselev}%
  \BibitemOpen
  \bibfield  {author} {\bibinfo {author} {\bibfnamefont {A.}~\bibnamefont
  {Kiselev}}, \bibinfo {author} {\bibfnamefont {M.}~\bibnamefont {Engel}}, \
  and\ \bibinfo {author} {\bibfnamefont {H.-R.}\ \bibnamefont {Trebin}},\
  }\href {\doibase 10.1103/PhysRevLett.109.225502} {\bibfield  {journal}
  {\bibinfo  {journal} {Phys. Rev. Lett.}\ }\textbf {\bibinfo {volume} {109}},\
  \bibinfo {pages} {225502} (\bibinfo {year} {2012})}\BibitemShut {NoStop}%
\bibitem [{\citenamefont {Haji-Akbari}\ \emph {et~al.}(2011)\citenamefont
  {Haji-Akbari}, \citenamefont {Engel},\ and\ \citenamefont
  {Glotzer}}]{Haji2011}%
  \BibitemOpen
  \bibfield  {author} {\bibinfo {author} {\bibfnamefont {A.}~\bibnamefont
  {Haji-Akbari}}, \bibinfo {author} {\bibfnamefont {M.}~\bibnamefont {Engel}},
  \ and\ \bibinfo {author} {\bibfnamefont {S.~C.}\ \bibnamefont {Glotzer}},\
  }\href {\doibase 10.1103/PhysRevLett.107.215702} {\bibfield  {journal}
  {\bibinfo  {journal} {Phys. Rev. Lett.}\ }\textbf {\bibinfo {volume} {107}},\
  \bibinfo {pages} {215702} (\bibinfo {year} {2011})}\BibitemShut {NoStop}%
\bibitem [{\citenamefont {Noya}\ \emph {et~al.}(2021)\citenamefont {Noya},
  \citenamefont {Wong}, \citenamefont {Llombart},\ and\ \citenamefont
  {Doye}}]{Doye}%
  \BibitemOpen
  \bibfield  {author} {\bibinfo {author} {\bibfnamefont {E.}~\bibnamefont
  {Noya}}, \bibinfo {author} {\bibfnamefont {C.~K.}\ \bibnamefont {Wong}},
  \bibinfo {author} {\bibfnamefont {P.}~\bibnamefont {Llombart}}, \ and\
  \bibinfo {author} {\bibfnamefont {J.~P.~K.}\ \bibnamefont {Doye}},\
  }\href@noop {} {\bibfield  {journal} {\bibinfo  {journal} {Nature}\ }\textbf
  {\bibinfo {volume} {596}},\ \bibinfo {pages} {367} (\bibinfo {year}
  {2021})}\BibitemShut {NoStop}%
\bibitem [{\citenamefont {Liu}\ \emph {et~al.}(2019)\citenamefont {Liu},
  \citenamefont {Li}, \citenamefont {Li},\ and\ \citenamefont {Mao}}]{DNA}%
  \BibitemOpen
  \bibfield  {author} {\bibinfo {author} {\bibfnamefont {L.}~\bibnamefont
  {Liu}}, \bibinfo {author} {\bibfnamefont {Z.}~\bibnamefont {Li}}, \bibinfo
  {author} {\bibfnamefont {Y.}~\bibnamefont {Li}}, \ and\ \bibinfo {author}
  {\bibfnamefont {C.}~\bibnamefont {Mao}},\ }\href@noop {} {\bibfield
  {journal} {\bibinfo  {journal} {J. Am. Chem. Soc.}\ }\textbf {\bibinfo
  {volume} {141}},\ \bibinfo {pages} {4248} (\bibinfo {year}
  {2019})}\BibitemShut {NoStop}%
\bibitem [{\citenamefont {Edwards}\ and\ \citenamefont
  {Fauve}(1994)}]{edwards_fauve_1994}%
  \BibitemOpen
  \bibfield  {author} {\bibinfo {author} {\bibfnamefont {W.~S.}\ \bibnamefont
  {Edwards}}\ and\ \bibinfo {author} {\bibfnamefont {S.}~\bibnamefont
  {Fauve}},\ }\href {\doibase 10.1017/S0022112094003642} {\bibfield  {journal}
  {\bibinfo  {journal} {J. Fluid Mech.}\ }\textbf {\bibinfo {volume} {278}},\
  \bibinfo {pages} {123–148} (\bibinfo {year} {1994})}\BibitemShut {NoStop}%
\bibitem [{\citenamefont {Lifshitz}\ and\ \citenamefont
  {Petrich}(1997)}]{LifshitzPetrich}%
  \BibitemOpen
  \bibfield  {author} {\bibinfo {author} {\bibfnamefont {R.}~\bibnamefont
  {Lifshitz}}\ and\ \bibinfo {author} {\bibfnamefont {D.~M.}\ \bibnamefont
  {Petrich}},\ }\href@noop {} {\bibfield  {journal} {\bibinfo  {journal} {Phys.
  Rev. Lett.}\ }\textbf {\bibinfo {volume} {79}},\ \bibinfo {pages} {1261}
  (\bibinfo {year} {1997})}\BibitemShut {NoStop}%
\bibitem [{\citenamefont {Ratliff}\ \emph {et~al.}(2019)\citenamefont
  {Ratliff}, \citenamefont {Archer}, \citenamefont {Subramanian},\ and\
  \citenamefont {Rucklidge}}]{Rutledge}%
  \BibitemOpen
  \bibfield  {author} {\bibinfo {author} {\bibfnamefont {D.~J.}\ \bibnamefont
  {Ratliff}}, \bibinfo {author} {\bibfnamefont {A.~J.}\ \bibnamefont {Archer}},
  \bibinfo {author} {\bibfnamefont {P.}~\bibnamefont {Subramanian}}, \ and\
  \bibinfo {author} {\bibfnamefont {A.~M.}\ \bibnamefont {Rucklidge}},\ }\href
  {\doibase 10.1103/PhysRevLett.123.148004} {\bibfield  {journal} {\bibinfo
  {journal} {Phys. Rev. Lett.}\ }\textbf {\bibinfo {volume} {123}},\ \bibinfo
  {pages} {148004} (\bibinfo {year} {2019})}\BibitemShut {NoStop}%
\bibitem [{\citenamefont {Macé}\ \emph {et~al.}(2016)\citenamefont {Macé},
  \citenamefont {Jagannathan},\ and\ \citenamefont {Duneau}}]{Mace}%
  \BibitemOpen
  \bibfield  {author} {\bibinfo {author} {\bibfnamefont {N.}~\bibnamefont
  {Macé}}, \bibinfo {author} {\bibfnamefont {A.}~\bibnamefont {Jagannathan}},
  \ and\ \bibinfo {author} {\bibfnamefont {M.}~\bibnamefont {Duneau}},\
  }\href@noop {} {\bibfield  {journal} {\bibinfo  {journal} {Crystals}\
  }\textbf {\bibinfo {volume} {6}} (\bibinfo {year} {2016})}\BibitemShut
  {NoStop}%
\bibitem [{\citenamefont {Viebahn}\ \emph {et~al.}(2019)\citenamefont
  {Viebahn}, \citenamefont {Sbroscia}, \citenamefont {Carter}, \citenamefont
  {Yu},\ and\ \citenamefont {Schneider}}]{ColdAtoms}%
  \BibitemOpen
  \bibfield  {author} {\bibinfo {author} {\bibfnamefont {K.}~\bibnamefont
  {Viebahn}}, \bibinfo {author} {\bibfnamefont {M.}~\bibnamefont {Sbroscia}},
  \bibinfo {author} {\bibfnamefont {E.}~\bibnamefont {Carter}}, \bibinfo
  {author} {\bibfnamefont {J.-C.}\ \bibnamefont {Yu}}, \ and\ \bibinfo {author}
  {\bibfnamefont {U.}~\bibnamefont {Schneider}},\ }\href {\doibase
  10.1103/PhysRevLett.122.110404} {\bibfield  {journal} {\bibinfo  {journal}
  {Phys. Rev. Lett.}\ }\textbf {\bibinfo {volume} {122}},\ \bibinfo {pages}
  {110404} (\bibinfo {year} {2019})}\BibitemShut {NoStop}%
\bibitem [{\citenamefont {Uri}\ \emph {et~al.}(2023)\citenamefont {Uri},
  \citenamefont {de~la Barrera}, \citenamefont {Randeria}, \citenamefont
  {Rodan-Legrain}, \citenamefont {Devakul}, \citenamefont {Crowley},
  \citenamefont {Paul}, \citenamefont {Watanabe}, \citenamefont {Taniguchi},
  \citenamefont {Lifshitz}, \citenamefont {Fu}, \citenamefont {Ashoori},\ and\
  \citenamefont {Jarillo-Herrero}}]{delaBarrera}%
  \BibitemOpen
  \bibfield  {author} {\bibinfo {author} {\bibfnamefont {A.}~\bibnamefont
  {Uri}}, \bibinfo {author} {\bibfnamefont {S.~C.}\ \bibnamefont {de~la
  Barrera}}, \bibinfo {author} {\bibfnamefont {M.~T.}\ \bibnamefont
  {Randeria}}, \bibinfo {author} {\bibfnamefont {D.}~\bibnamefont
  {Rodan-Legrain}}, \bibinfo {author} {\bibfnamefont {T.}~\bibnamefont
  {Devakul}}, \bibinfo {author} {\bibfnamefont {P.~J.~D.}\ \bibnamefont
  {Crowley}}, \bibinfo {author} {\bibfnamefont {N.}~\bibnamefont {Paul}},
  \bibinfo {author} {\bibfnamefont {K.}~\bibnamefont {Watanabe}}, \bibinfo
  {author} {\bibfnamefont {T.}~\bibnamefont {Taniguchi}}, \bibinfo {author}
  {\bibfnamefont {R.}~\bibnamefont {Lifshitz}}, \bibinfo {author}
  {\bibfnamefont {L.}~\bibnamefont {Fu}}, \bibinfo {author} {\bibfnamefont
  {R.~C.}\ \bibnamefont {Ashoori}}, \ and\ \bibinfo {author} {\bibfnamefont
  {P.}~\bibnamefont {Jarillo-Herrero}},\ }\href@noop {} {\enquote {\bibinfo
  {title} {Superconductivity and strong interactions in a tunable moir\'e
  quasiperiodic crystal},}\ } (\bibinfo {year} {2023}),\ \Eprint
  {http://arxiv.org/abs/2302.00686} {arXiv:2302.00686 [cond-mat.mes-hall]}
  \BibitemShut {NoStop}%
\bibitem [{\citenamefont {Shechtman}\ and\ \citenamefont
  {Blech}(1985)}]{ShechtmanBlech85}%
  \BibitemOpen
  \bibfield  {author} {\bibinfo {author} {\bibfnamefont {D.}~\bibnamefont
  {Shechtman}}\ and\ \bibinfo {author} {\bibfnamefont {I.}~\bibnamefont
  {Blech}},\ }\href@noop {} {\bibfield  {journal} {\bibinfo  {journal} {Metall.
  Trans. A}\ } (\bibinfo {year} {1985})}\BibitemShut {NoStop}%
\bibitem [{\citenamefont {Stephens}\ and\ \citenamefont
  {Goldman}(1986)}]{StephensGoldman}%
  \BibitemOpen
  \bibfield  {author} {\bibinfo {author} {\bibfnamefont {P.~W.}\ \bibnamefont
  {Stephens}}\ and\ \bibinfo {author} {\bibfnamefont {A.~I.}\ \bibnamefont
  {Goldman}},\ }\href@noop {} {\bibfield  {journal} {\bibinfo  {journal} {Phys.
  Rev. Lett.}\ }\textbf {\bibinfo {volume} {57}},\ \bibinfo {pages} {2331}
  (\bibinfo {year} {1986})}\BibitemShut {NoStop}%
\bibitem [{\citenamefont {Pauling}(1985)}]{Pauling85}%
  \BibitemOpen
  \bibfield  {author} {\bibinfo {author} {\bibfnamefont {L.}~\bibnamefont
  {Pauling}},\ }\href@noop {} {\bibfield  {journal} {\bibinfo  {journal}
  {Nature}\ }\textbf {\bibinfo {volume} {317}},\ \bibinfo {pages} {512}
  (\bibinfo {year} {1985})}\BibitemShut {NoStop}%
\bibitem [{\citenamefont {Pauling}(1987)}]{Pauling87}%
  \BibitemOpen
  \bibfield  {author} {\bibinfo {author} {\bibfnamefont {L.}~\bibnamefont
  {Pauling}},\ }\href@noop {} {\bibfield  {journal} {\bibinfo  {journal} {Phys.
  Rev. Lett.}\ }\textbf {\bibinfo {volume} {58}},\ \bibinfo {pages} {365}
  (\bibinfo {year} {1987})}\BibitemShut {NoStop}%
\bibitem [{\citenamefont {Takakura}\ \emph {et~al.}(2007)\citenamefont
  {Takakura}, \citenamefont {Gomez}, \citenamefont {Yamamoto}, \citenamefont
  {De~Boissieu},\ and\ \citenamefont {Tsai}}]{takakura-cdyb}%
  \BibitemOpen
  \bibfield  {author} {\bibinfo {author} {\bibfnamefont {H.}~\bibnamefont
  {Takakura}}, \bibinfo {author} {\bibfnamefont {C.~P.}\ \bibnamefont {Gomez}},
  \bibinfo {author} {\bibfnamefont {A.}~\bibnamefont {Yamamoto}}, \bibinfo
  {author} {\bibfnamefont {M.}~\bibnamefont {De~Boissieu}}, \ and\ \bibinfo
  {author} {\bibfnamefont {A.~P.}\ \bibnamefont {Tsai}},\ }\href@noop {}
  {\bibfield  {journal} {\bibinfo  {journal} {Nat. Mat.}\ }\textbf {\bibinfo
  {volume} {6}},\ \bibinfo {pages} {58} (\bibinfo {year} {2007})}\BibitemShut
  {NoStop}%
\bibitem [{\citenamefont {Katz}\ and\ \citenamefont {Gratias}(1993)}]{kg93}%
  \BibitemOpen
  \bibfield  {author} {\bibinfo {author} {\bibfnamefont {A.}~\bibnamefont
  {Katz}}\ and\ \bibinfo {author} {\bibfnamefont {D.}~\bibnamefont {Gratias}},\
  }\href@noop {} {\bibfield  {journal} {\bibinfo  {journal} {J. non-cryst.
  sol.}\ }\textbf {\bibinfo {volume} {153}},\ \bibinfo {pages} {187} (\bibinfo
  {year} {1993})}\BibitemShut {NoStop}%
\bibitem [{\citenamefont {Quiquandon}\ and\ \citenamefont
  {Gratias}(2006)}]{quiqua2006}%
  \BibitemOpen
  \bibfield  {author} {\bibinfo {author} {\bibfnamefont {M.}~\bibnamefont
  {Quiquandon}}\ and\ \bibinfo {author} {\bibfnamefont {D.}~\bibnamefont
  {Gratias}},\ }\href@noop {} {\bibfield  {journal} {\bibinfo  {journal} {Phys.
  Rev. B}\ }\textbf {\bibinfo {volume} {74}},\ \bibinfo {pages} {214205}
  (\bibinfo {year} {2006})}\BibitemShut {NoStop}%
\bibitem [{\citenamefont {Mihalkovi\ifmmode~\check{c}\else \v{c}\fi{}}\ and\
  \citenamefont {Widom}(2020)}]{i-AlCuFe}%
  \BibitemOpen
  \bibfield  {author} {\bibinfo {author} {\bibfnamefont {M.}~\bibnamefont
  {Mihalkovi\ifmmode~\check{c}\else \v{c}\fi{}}}\ and\ \bibinfo {author}
  {\bibfnamefont {M.}~\bibnamefont {Widom}},\ }\href {\doibase
  10.1103/PhysRevResearch.2.013196} {\bibfield  {journal} {\bibinfo  {journal}
  {Phys. Rev. Res.}\ }\textbf {\bibinfo {volume} {2}},\ \bibinfo {pages}
  {013196} (\bibinfo {year} {2020})}\BibitemShut {NoStop}%
\bibitem [{\citenamefont {Hohenberg}\ and\ \citenamefont
  {Kohn}(1964)}]{HohenbergKohn}%
  \BibitemOpen
  \bibfield  {author} {\bibinfo {author} {\bibfnamefont {P.}~\bibnamefont
  {Hohenberg}}\ and\ \bibinfo {author} {\bibfnamefont {W.}~\bibnamefont
  {Kohn}},\ }\href@noop {} {\bibfield  {journal} {\bibinfo  {journal} {Phys.
  Rev.}\ }\textbf {\bibinfo {volume} {136}},\ \bibinfo {pages} {B864} (\bibinfo
  {year} {1964})}\BibitemShut {NoStop}%
\bibitem [{\citenamefont {Kohn}\ and\ \citenamefont {Sham}(1965)}]{KohnSham}%
  \BibitemOpen
  \bibfield  {author} {\bibinfo {author} {\bibfnamefont {W.}~\bibnamefont
  {Kohn}}\ and\ \bibinfo {author} {\bibfnamefont {L.~J.}\ \bibnamefont
  {Sham}},\ }\href@noop {} {\bibfield  {journal} {\bibinfo  {journal} {Phys.
  Rev.}\ }\textbf {\bibinfo {volume} {140}},\ \bibinfo {pages} {A1133}
  (\bibinfo {year} {1965})}\BibitemShut {NoStop}%
\bibitem [{\citenamefont {Behler}\ and\ \citenamefont
  {Parrinello}(2007)}]{Behler}%
  \BibitemOpen
  \bibfield  {author} {\bibinfo {author} {\bibfnamefont {J.}~\bibnamefont
  {Behler}}\ and\ \bibinfo {author} {\bibfnamefont {M.}~\bibnamefont
  {Parrinello}},\ }\href {\doibase 10.1103/PhysRevLett.98.146401} {\bibfield
  {journal} {\bibinfo  {journal} {Phys. Rev. Lett.}\ }\textbf {\bibinfo
  {volume} {98}},\ \bibinfo {pages} {146401} (\bibinfo {year}
  {2007})}\BibitemShut {NoStop}%
\bibitem [{\citenamefont {Jinnouchi}\ \emph {et~al.}(2019)\citenamefont
  {Jinnouchi}, \citenamefont {Lahnsteiner}, \citenamefont {Karsai},
  \citenamefont {Kresse},\ and\ \citenamefont {Bokdam}}]{VASP-ML}%
  \BibitemOpen
  \bibfield  {author} {\bibinfo {author} {\bibfnamefont {R.}~\bibnamefont
  {Jinnouchi}}, \bibinfo {author} {\bibfnamefont {J.}~\bibnamefont
  {Lahnsteiner}}, \bibinfo {author} {\bibfnamefont {F.}~\bibnamefont {Karsai}},
  \bibinfo {author} {\bibfnamefont {G.}~\bibnamefont {Kresse}}, \ and\ \bibinfo
  {author} {\bibfnamefont {M.}~\bibnamefont {Bokdam}},\ }\href {\doibase
  10.1103/PhysRevLett.122.225701} {\bibfield  {journal} {\bibinfo  {journal}
  {Phys. Rev. Lett.}\ }\textbf {\bibinfo {volume} {122}},\ \bibinfo {pages}
  {225701} (\bibinfo {year} {2019})}\BibitemShut {NoStop}%
\bibitem [{\citenamefont {Friedel}(1958)}]{Friedel1958}%
  \BibitemOpen
  \bibfield  {author} {\bibinfo {author} {\bibfnamefont {J.}~\bibnamefont
  {Friedel}},\ }\href {\doibase 10.1007/BF02751483} {\bibfield  {journal}
  {\bibinfo  {journal} {Nuovo Cimento Suppl.}\ }\textbf {\bibinfo {volume}
  {7}},\ \bibinfo {pages} {287} (\bibinfo {year} {1958})}\BibitemShut {NoStop}%
\bibitem [{\citenamefont {Ashcroft}\ and\ \citenamefont
  {Lekner}(1966)}]{AshcroftLekner}%
  \BibitemOpen
  \bibfield  {author} {\bibinfo {author} {\bibfnamefont {N.~W.}\ \bibnamefont
  {Ashcroft}}\ and\ \bibinfo {author} {\bibfnamefont {J.}~\bibnamefont
  {Lekner}},\ }\href {\doibase 10.1103/PhysRev.145.83} {\bibfield  {journal}
  {\bibinfo  {journal} {Phys. Rev.}\ }\textbf {\bibinfo {volume} {145}},\
  \bibinfo {pages} {83} (\bibinfo {year} {1966})}\BibitemShut {NoStop}%
\bibitem [{\citenamefont {Hafner}(1987)}]{HafnerHam}%
  \BibitemOpen
  \bibfield  {author} {\bibinfo {author} {\bibfnamefont {J.}~\bibnamefont
  {Hafner}},\ }\href@noop {} {\emph {\bibinfo {title} {From Hamiltonians to
  Phase Diagrams}}}\ (\bibinfo  {publisher} {Springer-Verlag},\ \bibinfo
  {address} {Berlin},\ \bibinfo {year} {1987})\BibitemShut {NoStop}%
\bibitem [{\citenamefont {Moriarty}(1977)}]{MoriartyGPT}%
  \BibitemOpen
  \bibfield  {author} {\bibinfo {author} {\bibfnamefont {J.~A.}\ \bibnamefont
  {Moriarty}},\ }\href@noop {} {\bibfield  {journal} {\bibinfo  {journal}
  {Phys. Rev. B}\ }\textbf {\bibinfo {volume} {16}},\ \bibinfo {pages} {2537}
  (\bibinfo {year} {1977})}\BibitemShut {NoStop}%
\bibitem [{\citenamefont {Phillips}\ and\ \citenamefont
  {Widom}(1993)}]{Phillips1993}%
  \BibitemOpen
  \bibfield  {author} {\bibinfo {author} {\bibfnamefont {R.}~\bibnamefont
  {Phillips}}\ and\ \bibinfo {author} {\bibfnamefont {M.}~\bibnamefont
  {Widom}},\ }\href {\doibase https://doi.org/10.1016/0022-3093(93)90386-C}
  {\bibfield  {journal} {\bibinfo  {journal} {J. Non-Cryst. Sol.}\ }\textbf
  {\bibinfo {volume} {153-154}},\ \bibinfo {pages} {416} (\bibinfo {year}
  {1993})}\BibitemShut {NoStop}%
\bibitem [{\citenamefont {Mihalkovi{\v{c}}}\ \emph
  {et~al.}(1996{\natexlab{a}})\citenamefont {Mihalkovi{\v{c}}}, \citenamefont
  {Zhu}, \citenamefont {Henley},\ and\ \citenamefont {Phillips}}]{almn2}%
  \BibitemOpen
  \bibfield  {author} {\bibinfo {author} {\bibfnamefont {M.}~\bibnamefont
  {Mihalkovi{\v{c}}}}, \bibinfo {author} {\bibfnamefont {W.-J.}\ \bibnamefont
  {Zhu}}, \bibinfo {author} {\bibfnamefont {C.}~\bibnamefont {Henley}}, \ and\
  \bibinfo {author} {\bibfnamefont {R.}~\bibnamefont {Phillips}},\ }\href@noop
  {} {\bibfield  {journal} {\bibinfo  {journal} {Phys. Rev. B}\ }\textbf
  {\bibinfo {volume} {53}},\ \bibinfo {pages} {9021} (\bibinfo {year}
  {1996}{\natexlab{a}})}\BibitemShut {NoStop}%
\bibitem [{\citenamefont {Moriarty}\ and\ \citenamefont
  {Widom}(1997)}]{MoriartyAlTM}%
  \BibitemOpen
  \bibfield  {author} {\bibinfo {author} {\bibfnamefont {J.~A.}\ \bibnamefont
  {Moriarty}}\ and\ \bibinfo {author} {\bibfnamefont {M.}~\bibnamefont
  {Widom}},\ }\href {\doibase 10.1103/PhysRevB.56.7905} {\bibfield  {journal}
  {\bibinfo  {journal} {Phys. Rev. B}\ }\textbf {\bibinfo {volume} {56}},\
  \bibinfo {pages} {7905} (\bibinfo {year} {1997})}\BibitemShut {NoStop}%
\bibitem [{\citenamefont {Widom}\ \emph {et~al.}(2000)\citenamefont {Widom},
  \citenamefont {Al-Lehyani},\ and\ \citenamefont
  {Moriarty}}]{MoriartyTernary}%
  \BibitemOpen
  \bibfield  {author} {\bibinfo {author} {\bibfnamefont {M.}~\bibnamefont
  {Widom}}, \bibinfo {author} {\bibfnamefont {I.}~\bibnamefont {Al-Lehyani}}, \
  and\ \bibinfo {author} {\bibfnamefont {J.~A.}\ \bibnamefont {Moriarty}},\
  }\href {\doibase 10.1103/PhysRevB.62.3648} {\bibfield  {journal} {\bibinfo
  {journal} {Phys. Rev. B}\ }\textbf {\bibinfo {volume} {62}},\ \bibinfo
  {pages} {3648} (\bibinfo {year} {2000})}\BibitemShut {NoStop}%
\bibitem [{\citenamefont {Mihalkovi\ifmmode~\check{c}\else \v{c}\fi{}}\ and\
  \citenamefont {Henley}(2012)}]{EOPP}%
  \BibitemOpen
  \bibfield  {author} {\bibinfo {author} {\bibfnamefont {M.}~\bibnamefont
  {Mihalkovi\ifmmode~\check{c}\else \v{c}\fi{}}}\ and\ \bibinfo {author}
  {\bibfnamefont {C.~L.}\ \bibnamefont {Henley}},\ }\href {\doibase
  10.1103/PhysRevB.85.092102} {\bibfield  {journal} {\bibinfo  {journal} {Phys.
  Rev. B}\ }\textbf {\bibinfo {volume} {85}},\ \bibinfo {pages} {092102}
  (\bibinfo {year} {2012})}\BibitemShut {NoStop}%
\bibitem [{\citenamefont {Mihalkovi\ifmmode~\check{c}\else \v{c}\fi{}}\ and\
  \citenamefont {Widom}(2023)}]{i-AlMnPd}%
  \BibitemOpen
  \bibfield  {author} {\bibinfo {author} {\bibfnamefont {M.}~\bibnamefont
  {Mihalkovi\ifmmode~\check{c}\else \v{c}\fi{}}}\ and\ \bibinfo {author}
  {\bibfnamefont {M.}~\bibnamefont {Widom}},\ }\href@noop {} {\bibfield
  {journal} {\bibinfo  {journal} {Rend. Fis. Acc. Lincei}\ } (\bibinfo {year}
  {2023})}\BibitemShut {NoStop}%
\bibitem [{\citenamefont {Ishimasa}\ \emph {et~al.}(2011)\citenamefont
  {Ishimasa}, \citenamefont {Hirao}, \citenamefont {Honma},\ and\ \citenamefont
  {Mihalkovi{\v{c}}}}]{alcusc-ishimasa}%
  \BibitemOpen
  \bibfield  {author} {\bibinfo {author} {\bibfnamefont {T.}~\bibnamefont
  {Ishimasa}}, \bibinfo {author} {\bibfnamefont {A.}~\bibnamefont {Hirao}},
  \bibinfo {author} {\bibfnamefont {T.}~\bibnamefont {Honma}}, \ and\ \bibinfo
  {author} {\bibfnamefont {M.}~\bibnamefont {Mihalkovi{\v{c}}}},\ }\href@noop
  {} {\bibfield  {journal} {\bibinfo  {journal} {Phil. Mag.}\ }\textbf
  {\bibinfo {volume} {91}},\ \bibinfo {pages} {2594} (\bibinfo {year}
  {2011})}\BibitemShut {NoStop}%
\bibitem [{\citenamefont {Mihalkovi{\v{c}}}\ \emph {et~al.}(2014)\citenamefont
  {Mihalkovi{\v{c}}}, \citenamefont {Richmond-Decker}, \citenamefont {Henley},\
  and\ \citenamefont {Oxborrow}}]{d-myz}%
  \BibitemOpen
  \bibfield  {author} {\bibinfo {author} {\bibfnamefont {M.}~\bibnamefont
  {Mihalkovi{\v{c}}}}, \bibinfo {author} {\bibfnamefont {J.}~\bibnamefont
  {Richmond-Decker}}, \bibinfo {author} {\bibfnamefont {C.}~\bibnamefont
  {Henley}}, \ and\ \bibinfo {author} {\bibfnamefont {M.}~\bibnamefont
  {Oxborrow}},\ }\href@noop {} {\bibfield  {journal} {\bibinfo  {journal}
  {Phil. Mag.}\ }\textbf {\bibinfo {volume} {94}},\ \bibinfo {pages} {1529}
  (\bibinfo {year} {2014})}\BibitemShut {NoStop}%
\bibitem [{\citenamefont {De~Boissieu}\ \emph {et~al.}(2007)\citenamefont
  {De~Boissieu}, \citenamefont {Francoual}, \citenamefont {Mihalkovi{\v{c}}},
  \citenamefont {Shibata}, \citenamefont {Baron}, \citenamefont {Sidis},
  \citenamefont {Ishimasa}, \citenamefont {Wu}, \citenamefont {Lograsso},
  \citenamefont {Regnault} \emph {et~al.}}]{sczn-phonon}%
  \BibitemOpen
  \bibfield  {author} {\bibinfo {author} {\bibfnamefont {M.}~\bibnamefont
  {De~Boissieu}}, \bibinfo {author} {\bibfnamefont {S.}~\bibnamefont
  {Francoual}}, \bibinfo {author} {\bibfnamefont {M.}~\bibnamefont
  {Mihalkovi{\v{c}}}}, \bibinfo {author} {\bibfnamefont {K.}~\bibnamefont
  {Shibata}}, \bibinfo {author} {\bibfnamefont {A.~Q.}\ \bibnamefont {Baron}},
  \bibinfo {author} {\bibfnamefont {Y.}~\bibnamefont {Sidis}}, \bibinfo
  {author} {\bibfnamefont {T.}~\bibnamefont {Ishimasa}}, \bibinfo {author}
  {\bibfnamefont {D.}~\bibnamefont {Wu}}, \bibinfo {author} {\bibfnamefont
  {T.}~\bibnamefont {Lograsso}}, \bibinfo {author} {\bibfnamefont {L.-P.}\
  \bibnamefont {Regnault}},  \emph {et~al.},\ }\href@noop {} {\bibfield
  {journal} {\bibinfo  {journal} {Nat. Mat.}\ }\textbf {\bibinfo {volume}
  {6}},\ \bibinfo {pages} {977} (\bibinfo {year} {2007})}\BibitemShut {NoStop}%
\bibitem [{\citenamefont {Mihalkovi{\v{c}}}\ and\ \citenamefont
  {Henley}(2011)}]{sczn-tet}%
  \BibitemOpen
  \bibfield  {author} {\bibinfo {author} {\bibfnamefont {M.}~\bibnamefont
  {Mihalkovi{\v{c}}}}\ and\ \bibinfo {author} {\bibfnamefont {C.~L.}\
  \bibnamefont {Henley}},\ }\href@noop {} {\bibfield  {journal} {\bibinfo
  {journal} {Phil. Mag.}\ }\textbf {\bibinfo {volume} {91}},\ \bibinfo {pages}
  {2548} (\bibinfo {year} {2011})}\BibitemShut {NoStop}%
\bibitem [{\citenamefont {Feuerbacher}\ \emph {et~al.}(2007)\citenamefont
  {Feuerbacher}, \citenamefont {Thomas}, \citenamefont {Makongo}, \citenamefont
  {Hoffmann}, \citenamefont {Carrillo-Cabrera}, \citenamefont {Cardoso},
  \citenamefont {Grin}, \citenamefont {Kreiner}, \citenamefont {Joubert},
  \citenamefont {Schenk} \emph {et~al.}}]{al3mg2}%
  \BibitemOpen
  \bibfield  {author} {\bibinfo {author} {\bibfnamefont {M.}~\bibnamefont
  {Feuerbacher}}, \bibinfo {author} {\bibfnamefont {C.}~\bibnamefont {Thomas}},
  \bibinfo {author} {\bibfnamefont {J.~P.}\ \bibnamefont {Makongo}}, \bibinfo
  {author} {\bibfnamefont {S.}~\bibnamefont {Hoffmann}}, \bibinfo {author}
  {\bibfnamefont {W.}~\bibnamefont {Carrillo-Cabrera}}, \bibinfo {author}
  {\bibfnamefont {R.}~\bibnamefont {Cardoso}}, \bibinfo {author} {\bibfnamefont
  {Y.}~\bibnamefont {Grin}}, \bibinfo {author} {\bibfnamefont {G.}~\bibnamefont
  {Kreiner}}, \bibinfo {author} {\bibfnamefont {J.-M.}\ \bibnamefont
  {Joubert}}, \bibinfo {author} {\bibfnamefont {T.}~\bibnamefont {Schenk}},
  \emph {et~al.},\ }\href@noop {} {\bibfield  {journal} {\bibinfo  {journal}
  {Zeit. Krist.}\ }\textbf {\bibinfo {volume} {222}},\ \bibinfo {pages} {259}
  (\bibinfo {year} {2007})}\BibitemShut {NoStop}%
\bibitem [{\citenamefont {Mihalkovi{\v{c}}}\ and\ \citenamefont
  {Henley}(2013)}]{al11ir4}%
  \BibitemOpen
  \bibfield  {author} {\bibinfo {author} {\bibfnamefont {M.}~\bibnamefont
  {Mihalkovi{\v{c}}}}\ and\ \bibinfo {author} {\bibfnamefont {C.}~\bibnamefont
  {Henley}},\ }\href@noop {} {\bibfield  {journal} {\bibinfo  {journal} {Phys.
  Rev. B}\ }\textbf {\bibinfo {volume} {88}},\ \bibinfo {pages} {064201}
  (\bibinfo {year} {2013})}\BibitemShut {NoStop}%
\bibitem [{\citenamefont {Mihalkovi\v{c}}\ \emph {et~al.}(2011)\citenamefont
  {Mihalkovi\v{c}}, \citenamefont {Widom},\ and\ \citenamefont
  {Henley}}]{CellConstraint}%
  \BibitemOpen
  \bibfield  {author} {\bibinfo {author} {\bibfnamefont {M.}~\bibnamefont
  {Mihalkovi\v{c}}}, \bibinfo {author} {\bibfnamefont {M.}~\bibnamefont
  {Widom}}, \ and\ \bibinfo {author} {\bibfnamefont {C.~L.}\ \bibnamefont
  {Henley}},\ }\href@noop {} {\bibfield  {journal} {\bibinfo  {journal} {Phil.
  Mag.}\ }\textbf {\bibinfo {volume} {91}},\ \bibinfo {pages} {2557} (\bibinfo
  {year} {2011})}\BibitemShut {NoStop}%
\bibitem [{\citenamefont {Widom}(2008)}]{PhasonReview}%
  \BibitemOpen
  \bibfield  {author} {\bibinfo {author} {\bibfnamefont {M.}~\bibnamefont
  {Widom}},\ }\href@noop {} {\bibfield  {journal} {\bibinfo  {journal} {Phil.
  Mag.}\ }\textbf {\bibinfo {volume} {88}},\ \bibinfo {pages} {2339} (\bibinfo
  {year} {2008})}\BibitemShut {NoStop}%
\bibitem [{\citenamefont {Naidu}\ \emph {et~al.}(2005)\citenamefont {Naidu},
  \citenamefont {Mihalkovi\v{c}},\ and\ \citenamefont {Widom}}]{Naidu}%
  \BibitemOpen
  \bibfield  {author} {\bibinfo {author} {\bibfnamefont {S.}~\bibnamefont
  {Naidu}}, \bibinfo {author} {\bibfnamefont {M.}~\bibnamefont
  {Mihalkovi\v{c}}}, \ and\ \bibinfo {author} {\bibfnamefont {M.}~\bibnamefont
  {Widom}},\ }\href@noop {} {\bibfield  {journal} {\bibinfo  {journal} {Phys.
  Rev. B}\ }\textbf {\bibinfo {volume} {71}},\ \bibinfo {pages} {224207}
  (\bibinfo {year} {2005})}\BibitemShut {NoStop}%
\bibitem [{\citenamefont {Bak}(1985)}]{Bak1985}%
  \BibitemOpen
  \bibfield  {author} {\bibinfo {author} {\bibfnamefont {P.}~\bibnamefont
  {Bak}},\ }\href@noop {} {\bibfield  {journal} {\bibinfo  {journal} {Phys.
  Rev. B}\ }\textbf {\bibinfo {volume} {32}},\ \bibinfo {pages} {5764}
  (\bibinfo {year} {1985})}\BibitemShut {NoStop}%
\bibitem [{\citenamefont {Lubensky}\ \emph {et~al.}(1985)\citenamefont
  {Lubensky}, \citenamefont {Ramaswamy},\ and\ \citenamefont
  {Toner}}]{Lubensky1985}%
  \BibitemOpen
  \bibfield  {author} {\bibinfo {author} {\bibfnamefont {T.~C.}\ \bibnamefont
  {Lubensky}}, \bibinfo {author} {\bibfnamefont {S.}~\bibnamefont {Ramaswamy}},
  \ and\ \bibinfo {author} {\bibfnamefont {J.}~\bibnamefont {Toner}},\
  }\href@noop {} {\bibfield  {journal} {\bibinfo  {journal} {Phys. Rev. B}\
  }\textbf {\bibinfo {volume} {32}},\ \bibinfo {pages} {7444} (\bibinfo {year}
  {1985})}\BibitemShut {NoStop}%
\bibitem [{\citenamefont {Mihalkovi{\v{c}}}\ \emph
  {et~al.}(1996{\natexlab{b}})\citenamefont {Mihalkovi{\v{c}}}, \citenamefont
  {Zhu}, \citenamefont {Henley},\ and\ \citenamefont {Oxborrow}}]{almn1}%
  \BibitemOpen
  \bibfield  {author} {\bibinfo {author} {\bibfnamefont {M.}~\bibnamefont
  {Mihalkovi{\v{c}}}}, \bibinfo {author} {\bibfnamefont {W.-J.}\ \bibnamefont
  {Zhu}}, \bibinfo {author} {\bibfnamefont {C.}~\bibnamefont {Henley}}, \ and\
  \bibinfo {author} {\bibfnamefont {M.}~\bibnamefont {Oxborrow}},\ }\href@noop
  {} {\bibfield  {journal} {\bibinfo  {journal} {Physical Review B}\ }\textbf
  {\bibinfo {volume} {53}},\ \bibinfo {pages} {9002} (\bibinfo {year}
  {1996}{\natexlab{b}})}\BibitemShut {NoStop}%
\bibitem [{\citenamefont {Mihalkovi\ifmmode~\check{c}\else \v{c}\fi{}}\ \emph
  {et~al.}(2002)\citenamefont {Mihalkovi\ifmmode~\check{c}\else \v{c}\fi{}},
  \citenamefont {Al-Lehyani}, \citenamefont {Cockayne}, \citenamefont {Henley},
  \citenamefont {Moghadam}, \citenamefont {Moriarty}, \citenamefont {Wang},\
  and\ \citenamefont {Widom}}]{AlCoNi-deco}%
  \BibitemOpen
  \bibfield  {author} {\bibinfo {author} {\bibfnamefont {M.}~\bibnamefont
  {Mihalkovi\ifmmode~\check{c}\else \v{c}\fi{}}}, \bibinfo {author}
  {\bibfnamefont {I.}~\bibnamefont {Al-Lehyani}}, \bibinfo {author}
  {\bibfnamefont {E.}~\bibnamefont {Cockayne}}, \bibinfo {author}
  {\bibfnamefont {C.~L.}\ \bibnamefont {Henley}}, \bibinfo {author}
  {\bibfnamefont {N.}~\bibnamefont {Moghadam}}, \bibinfo {author}
  {\bibfnamefont {J.~A.}\ \bibnamefont {Moriarty}}, \bibinfo {author}
  {\bibfnamefont {Y.}~\bibnamefont {Wang}}, \ and\ \bibinfo {author}
  {\bibfnamefont {M.}~\bibnamefont {Widom}},\ }\href {\doibase
  10.1103/PhysRevB.65.104205} {\bibfield  {journal} {\bibinfo  {journal} {Phys.
  Rev. B}\ }\textbf {\bibinfo {volume} {65}},\ \bibinfo {pages} {104205}
  (\bibinfo {year} {2002})}\BibitemShut {NoStop}%
\bibitem [{\citenamefont {Henley}\ \emph {et~al.}(2002)\citenamefont {Henley},
  \citenamefont {Mihalkovi\v{c}},\ and\ \citenamefont {Widom}}]{T-Ham-AlNiCo}%
  \BibitemOpen
  \bibfield  {author} {\bibinfo {author} {\bibfnamefont {C.}~\bibnamefont
  {Henley}}, \bibinfo {author} {\bibfnamefont {M.}~\bibnamefont
  {Mihalkovi\v{c}}}, \ and\ \bibinfo {author} {\bibfnamefont {M.}~\bibnamefont
  {Widom}},\ }\href {\doibase https://doi.org/10.1016/S0925-8388(02)00199-8}
  {\bibfield  {journal} {\bibinfo  {journal} {J. Alloys Compd.}\ }\textbf
  {\bibinfo {volume} {342}},\ \bibinfo {pages} {221} (\bibinfo {year}
  {2002})}\BibitemShut {NoStop}%
\bibitem [{\citenamefont {Al-Lehyani}\ and\ \citenamefont
  {Widom}(2003)}]{T-Ham-AlCoCu-FP}%
  \BibitemOpen
  \bibfield  {author} {\bibinfo {author} {\bibfnamefont {I.}~\bibnamefont
  {Al-Lehyani}}\ and\ \bibinfo {author} {\bibfnamefont {M.}~\bibnamefont
  {Widom}},\ }\href {\doibase 10.1103/PhysRevB.67.014204} {\bibfield  {journal}
  {\bibinfo  {journal} {Phys. Rev. B}\ }\textbf {\bibinfo {volume} {67}},\
  \bibinfo {pages} {014204} (\bibinfo {year} {2003})}\BibitemShut {NoStop}%
\bibitem [{\citenamefont {Widom}\ \emph {et~al.}(2004)\citenamefont {Widom},
  \citenamefont {Al-Lehyani},\ and\ \citenamefont
  {Mihalkovi\v{c}}}]{T-Ham-AlCoCu}%
  \BibitemOpen
  \bibfield  {author} {\bibinfo {author} {\bibfnamefont {M.}~\bibnamefont
  {Widom}}, \bibinfo {author} {\bibfnamefont {I.}~\bibnamefont {Al-Lehyani}}, \
  and\ \bibinfo {author} {\bibfnamefont {M.}~\bibnamefont {Mihalkovi\v{c}}},\
  }\href {\doibase https://doi.org/10.1016/j.jnoncrysol.2003.11.018} {\bibfield
   {journal} {\bibinfo  {journal} {J. Non-Crystalline Sol.}\ }\textbf {\bibinfo
  {volume} {334-335}},\ \bibinfo {pages} {86} (\bibinfo {year}
  {2004})}\BibitemShut {NoStop}%
\bibitem [{\citenamefont {Roth}\ and\ \citenamefont {Henley}(1997)}]{roth1997}%
  \BibitemOpen
  \bibfield  {author} {\bibinfo {author} {\bibfnamefont {J.}~\bibnamefont
  {Roth}}\ and\ \bibinfo {author} {\bibfnamefont {C.~L.}\ \bibnamefont
  {Henley}},\ }\href@noop {} {\bibfield  {journal} {\bibinfo  {journal} {Phil.
  Mag. A}\ }\textbf {\bibinfo {volume} {75}},\ \bibinfo {pages} {861} (\bibinfo
  {year} {1997})}\BibitemShut {NoStop}%
\bibitem [{\citenamefont {Swendsen}\ and\ \citenamefont
  {Wang}(1986)}]{SwendsenWang}%
  \BibitemOpen
  \bibfield  {author} {\bibinfo {author} {\bibfnamefont {R.~H.}\ \bibnamefont
  {Swendsen}}\ and\ \bibinfo {author} {\bibfnamefont {J.-S.}\ \bibnamefont
  {Wang}},\ }\href@noop {} {\bibfield  {journal} {\bibinfo  {journal} {Phys.
  Rev. Lett.}\ }\textbf {\bibinfo {volume} {57}},\ \bibinfo {pages} {2607}
  (\bibinfo {year} {1986})}\BibitemShut {NoStop}%
\bibitem [{\citenamefont {Kim}\ and\ \citenamefont {Widom}(2023)}]{KimWidom}%
  \BibitemOpen
  \bibfield  {author} {\bibinfo {author} {\bibfnamefont {A.~D.}\ \bibnamefont
  {Kim}}\ and\ \bibinfo {author} {\bibfnamefont {M.}~\bibnamefont {Widom}},\
  }\href@noop {} {\bibfield  {journal} {\bibinfo  {journal} {Phys. Rev.
  Mater.}\ }\textbf {\bibinfo {volume} {7}},\ \bibinfo {pages} {063803}
  (\bibinfo {year} {2023})}\BibitemShut {NoStop}%
\bibitem [{\citenamefont {Mihalkovi\v{c}}\ \emph {et~al.}(2004)\citenamefont
  {Mihalkovi\v{c}}, \citenamefont {Henley},\ and\ \citenamefont
  {Widom}}]{Combined}%
  \BibitemOpen
  \bibfield  {author} {\bibinfo {author} {\bibfnamefont {M.}~\bibnamefont
  {Mihalkovi\v{c}}}, \bibinfo {author} {\bibfnamefont {C.~L.}\ \bibnamefont
  {Henley}}, \ and\ \bibinfo {author} {\bibfnamefont {M.}~\bibnamefont
  {Widom}},\ }\href@noop {} {\bibfield  {journal} {\bibinfo  {journal} {J.
  Non-Cryst. Sol.}\ }\textbf {\bibinfo {volume} {334-335}},\ \bibinfo {pages}
  {177} (\bibinfo {year} {2004})}\BibitemShut {NoStop}%
\bibitem [{\citenamefont {Frenkel}\ and\ \citenamefont
  {Smit}(2001)}]{FrenkelSmit}%
  \BibitemOpen
  \bibfield  {author} {\bibinfo {author} {\bibfnamefont {D.}~\bibnamefont
  {Frenkel}}\ and\ \bibinfo {author} {\bibfnamefont {B.}~\bibnamefont {Smit}},\
  }\href@noop {} {\emph {\bibinfo {title} {Understanding Molecular Simulation:
  From Algorithms to Applications}}}\ (\bibinfo  {publisher} {Elsevier},\
  \bibinfo {year} {2001})\BibitemShut {NoStop}%
\bibitem [{\citenamefont {Reinhardt}\ \emph {et~al.}(2013)\citenamefont
  {Reinhardt}, \citenamefont {Romano},\ and\ \citenamefont {Doye}}]{Reinhardt}%
  \BibitemOpen
  \bibfield  {author} {\bibinfo {author} {\bibfnamefont {A.}~\bibnamefont
  {Reinhardt}}, \bibinfo {author} {\bibfnamefont {F.}~\bibnamefont {Romano}}, \
  and\ \bibinfo {author} {\bibfnamefont {J.~P.~K.}\ \bibnamefont {Doye}},\
  }\href {\doibase 10.1103/PhysRevLett.110.255503} {\bibfield  {journal}
  {\bibinfo  {journal} {Phys. Rev. Lett.}\ }\textbf {\bibinfo {volume} {110}},\
  \bibinfo {pages} {255503} (\bibinfo {year} {2013})}\BibitemShut {NoStop}%
\bibitem [{\citenamefont {Pattabhiraman}\ \emph {et~al.}(2015)\citenamefont
  {Pattabhiraman}, \citenamefont {Gantapara},\ and\ \citenamefont
  {Dijkstra}}]{Dijkstra}%
  \BibitemOpen
  \bibfield  {author} {\bibinfo {author} {\bibfnamefont {H.}~\bibnamefont
  {Pattabhiraman}}, \bibinfo {author} {\bibfnamefont {A.~P.}\ \bibnamefont
  {Gantapara}}, \ and\ \bibinfo {author} {\bibfnamefont {M.}~\bibnamefont
  {Dijkstra}},\ }\href@noop {} {\bibfield  {journal} {\bibinfo  {journal} {J.
  Chem. Phys.}\ }\textbf {\bibinfo {volume} {143}},\ \bibinfo {pages} {164905}
  (\bibinfo {year} {2015})}\BibitemShut {NoStop}%
\bibitem [{\citenamefont {Fultz}(2010)}]{Fultz2010}%
  \BibitemOpen
  \bibfield  {author} {\bibinfo {author} {\bibfnamefont {B.}~\bibnamefont
  {Fultz}},\ }\href@noop {} {\bibfield  {journal} {\bibinfo  {journal} {Prog.
  Mat. Sci.}\ }\textbf {\bibinfo {volume} {55}},\ \bibinfo {pages} {247}
  (\bibinfo {year} {2010})}\BibitemShut {NoStop}%
\bibitem [{\citenamefont {Wolverton}\ and\ \citenamefont {Ozoli\ifmmode
  \mbox{\c{n}}\else \c{n}\fi{}\ifmmode~\check{s}\else
  \v{s}\fi{}}(2001)}]{Wolverton2001}%
  \BibitemOpen
  \bibfield  {author} {\bibinfo {author} {\bibfnamefont {C.}~\bibnamefont
  {Wolverton}}\ and\ \bibinfo {author} {\bibfnamefont {V.}~\bibnamefont
  {Ozoli\ifmmode \mbox{\c{n}}\else \c{n}\fi{}\ifmmode~\check{s}\else
  \v{s}\fi{}}},\ }\href@noop {} {\bibfield  {journal} {\bibinfo  {journal}
  {Phys. Rev. Lett.}\ }\textbf {\bibinfo {volume} {86}},\ \bibinfo {pages}
  {5518} (\bibinfo {year} {2001})}\BibitemShut {NoStop}%
\bibitem [{\citenamefont {Huang}\ and\ \citenamefont {Widom}(2022)}]{Svib}%
  \BibitemOpen
  \bibfield  {author} {\bibinfo {author} {\bibfnamefont {Y.}~\bibnamefont
  {Huang}}\ and\ \bibinfo {author} {\bibfnamefont {M.}~\bibnamefont {Widom}},\
  }\href@noop {} {\bibfield  {journal} {\bibinfo  {journal} {Entropy}\ }\textbf
  {\bibinfo {volume} {24}},\ \bibinfo {pages} {618} (\bibinfo {year}
  {2022})}\BibitemShut {NoStop}%
\bibitem [{\citenamefont {Widom}(2015)}]{Chapter8}%
  \BibitemOpen
  \bibfield  {author} {\bibinfo {author} {\bibfnamefont {M.}~\bibnamefont
  {Widom}},\ }\enquote {\bibinfo {title} {Prediction of structure and phase
  transformations},}\ \ (\bibinfo  {publisher} {Springer},\ \bibinfo {year}
  {2015})\ Chap.\ \bibinfo {chapter} {8 in {\em High Entropy Alloys:
  fundamentals and applications}, eds. Gao, Yeh, Liaw and Zhang}\BibitemShut
  {NoStop}%
\end{thebibliography}%
\end{document}